\pdfoutput=1

\documentclass[review,3p,times]{elsarticle}

\usepackage{bm}
\usepackage{amsmath}
\usepackage{mathtools}
\usepackage{dirtytalk}
\usepackage{derivative}
\usepackage{threeparttable}
\usepackage{lscape}
\usepackage{float}
\usepackage{lineno}
\usepackage{booktabs}
\usepackage{threeparttable}
\usepackage{caption}
\usepackage{pgfplots}
\usepackage{subcaption}
\usepackage[short,nocomma]{optidef}
\usepackage{hyperref}
\usepackage{tikz}
\usepackage{tabularx}
\usepackage{array}
\usepackage{multirow}

\makeatletter
\def\ps@pprintTitle{%
  \let\@oddhead\@empty
  \let\@evenhead\@empty
  \def\@oddfoot{\reset@font\hfil\thepage\hfil}
  \let\@evenfoot\@oddfoot
}
\makeatother

\pgfplotsset{compat=1.18}
\begin{document}
% \linenumbers

\begin{frontmatter}

\author[1]{Sebastián Espinel-Ríos\corref{cor1}}
\ead{sebastian.espinelrios@ucd.ie}

\author[2]{Wenchao Duan\corref{cor1}}\ead{wenchao.duan@csiro.au}

\cortext[cor1]{Corresponding author}

\affiliation[1]{organization={Department of Chemical and Bioprocess Engineering, University College Dublin},
            city={Dublin},
            % postcode={}, 
            % state={},
            country={Ireland}}

\affiliation[2]{organization={Commonwealth Scientific and Industrial Research Organisation},
            city={Clayton},
            % postcode={}, 
            % state={},
            country={Australia}}

\title{Automated feature-region selection for soft-sensor development from spectral-like measurements}

\begin{abstract}
Sustainable production increasingly relies on process analytical chemistry and process analytical technology to support monitoring, control, and automation. Techniques used in these contexts, including Raman spectroscopy, infrared spectroscopy, and electrochemical voltammetry, generate high-dimensional signals ordered along physical measurement axes, referred to here as \textit{spectral-like measurements}. These signals can contain redundant, weakly informative, and noisy regions, complicating the development of data-driven soft sensors to map them to process variables. Selecting informative regions is nontrivial, as visually prominent regions are not necessarily the most predictive, while synergistic effects among regions cannot be readily inferred. Here, we introduce an automated feature-region selection framework that identifies a parsimonious set of contiguous regions while preserving channel ordering. The framework combines channel-level target correlation and a signal-to-noise indicator into a latent information fingerprint. Variable-width candidate intervals are derived from peaks in this fingerprint, and their combinations are ranked using held-out validation data. Final selection favours models using fewer channels among candidates with comparable predictive performance. The framework is demonstrated using cyclic voltammetric measurements of glucose acquired with a gold electrode sensor across four progressively broader nominal concentration ranges up to 350~g/L. Gaussian process regression is used within the framework to accommodate possible nonlinear relationships, account for observation noise, and quantify epistemic uncertainty. The selected models retained only 11--42 of the original 400 channels and reduced held-out test root mean squared error by 88.3--94.6\% relative to full-feature Gaussian process regression benchmarks.
\end{abstract}

\begin{graphicalabstract}
\centering
% \vspace*{-4cm}
\includegraphics[]{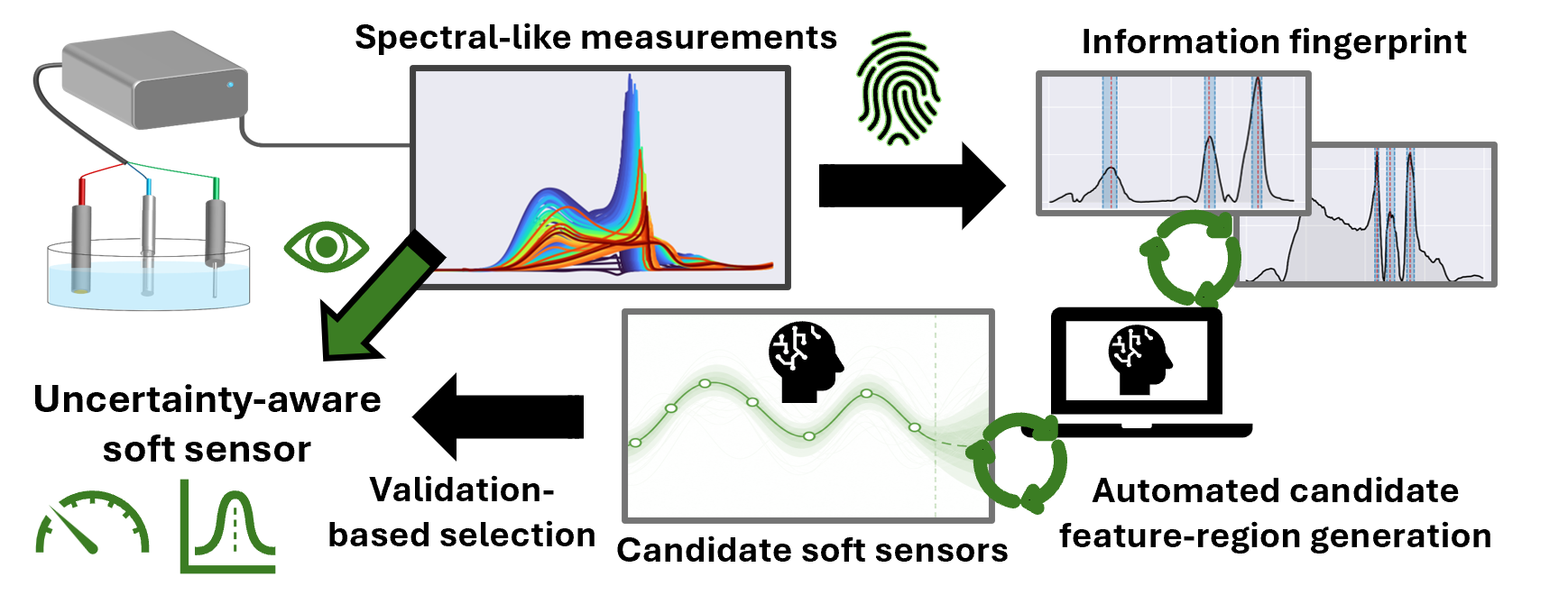}
\end{graphicalabstract}

\begin{highlights} 
\item soft sensors, automated feature selection, high dimension inputs, process monitoring, cyclic voltammogeam, glucose.
\end{highlights}

\begin{highlights}
\item Automated feature selection finds contiguous regions in spectral-like measurements.
\item A latent fingerprint combines target correlation and a signal-to-noise indicator.
\item Candidate intervals are derived from the fingerprint and evaluated in combination.
\item Gaussian-process-based glucose soft sensors use 11--42 of 400 voltammetric channels.
\item Test root mean squared error falls by 88.3--94.6\% versus full-feature models.
\end{highlights}


\end{frontmatter}

\section{Introduction}
Sustainable production systems are central to the transition towards a biobased and circular economy. Bioproduction and biorefinery systems contribute to this transition by enabling the manufacturing of chemicals, materials, and fuels from renewable feedstocks, including lignocellulosic and starch-rich biomass \cite{calvo-flores_biorefineries_2022}. In line with Industry 4.0 and Quality-by-Control (QbC), advanced process control approaches, including model predictive control and reinforcement learning, can improve the efficiency and robustness of chemical and biotechnological processes \cite{isoko_bioprocessing_2024, bloor_survey_2026}. Yet their implementation requires real-time information about relevant process variables to adjust control inputs in response to disturbances and uncertainty. Consequently, their feasibility depends on the availability of reliable online measurements or estimates of key variables, including the concentrations of substrates, reactants, intermediates, products, and by-products. This need has motivated advances in process analytical chemistry (PAC) and in process analytical technology (PAT) for process monitoring \cite{gerzon_process_2022, noauthor_multiway_2024}. Many analytical techniques based on hardware sensors generate
high-dimensional signals sampled along a physical measurement axis,
such as wavenumber or applied potential. Examples include Raman \cite{esmonde-white_role_2022} and infrared spectroscopy \cite{pu_review_2020}, as well as electrochemical voltammetry \cite{matosperalta_experimental_2026}. Here, we refer to measurements with this ordered signal structure as \textit{spectral-like measurements}.

For such multivariate signals, soft sensors bridge measurements acquired by hardware sensors and process variables or other quantities of interest that are not measured directly. They achieve this using mathematical models, typically calibrated using experimental data \cite{rathore_digitization_2022}. More broadly, state estimators can also perform soft-sensing tasks by using dynamic models to reconstruct system states from partial measurements \cite{mcafee_state_2022,elsheikh_comparative_2021}. In this work, we focus on calibration-based soft sensors. Among the widely adopted approaches, chemometric methods based on partial least-squares (PLS) regression \cite{wold_pls-regression_2001,durauer_sensors_2024} are particularly used for this purpose, as they accommodate high-dimensional, correlated
measurements through a reduced set of latent variables. However, the linear formulation of conventional PLS regression can limit its predictive performance when nonlinear relationships are pronounced. In such cases, nonlinear regression models, including parametric approaches such as neural networks and non-parametric approaches such as random forests and Gaussian processes (GPs), offer promising alternatives \cite{durauer_sensors_2024}. Nevertheless, when experimental datasets are limited, high input dimensionality and feature redundancy can increase data requirements and the risk of overfitting, while measurement noise can further complicate the development of data-driven soft sensors \cite{borg_less_2026,guo_semisupervised_2021,peng_machine_2026}.

Motivated by these challenges, we propose an automated framework for soft-sensor development through feature-region selection in spectral-like measurements. Its goal is to identify a compact set of informative signal regions, thereby reducing input dimensionality while retaining relevant information for soft-sensor development. To this end, the framework combines channel-level assessment of target correlation and measurement quality with model-based evaluation of candidate interval combinations. Using training data, the methodology first computes a pointwise score for each measurement channel by combining the square of its Spearman rank correlation coefficient with the target variable and a variance-based signal-to-noise indicator derived from replicate measurements. The resulting composite scores form a profile that reveals a latent information pattern along the measurement axis, herein referred to as the \emph{information fingerprint}. Promising feature intervals are automatically identified from peaks in this information fingerprint and prioritised according to the average scores of their constituent channels. Notably, each interval comprises contiguous measurement channels, preserving the local ordering of the signal along the physical measurement axis. Soft-sensor models are then trained using candidate combinations of these intervals and evaluated on held-out validation data. Final selection favours parsimony, i.e., models with fewer input channels among candidates with comparable validation performance.

The proposed framework avoids manual selection of peaks or intervals in spectral-like measurements, which can require application-specific domain knowledge and may overlook informative regions that are not immediately evident from visual inspection. Unlike conventional principal component analysis (PCA) \cite{peng_machine_2026}, which constructs uncorrelated latent components from linear combinations of the full input features by maximising explained input variance without using target information, our framework retains only a compact set of contiguous regions from the original signal. It does so using a latent information fingerprint that explicitly combines target correlation with replicate-based measurement quality, without requiring a linear relationship between individual channels and the target variable. Unlike standard permutation feature importance \cite{breiman_random_2001}, which evaluates feature contributions using an already fitted prediction model and does not inherently enforce contiguity along the measurement axis, our framework generates contiguous candidate regions from peaks in the information fingerprint before regression fitting. In conventional interval PLS \cite{norgaard_interval_2000}, predefined equal-width spectral intervals are successively added or removed according to cross-validated PLS prediction performance. Our candidate intervals can vary in width, and their generation is independent of the downstream regression architecture.

Although candidate interval generation is model-agnostic, Gaussian process regression (GPR) \cite{rasmussen_gaussian_2006} is used as the downstream modelling approach in this study. With an appropriate kernel function, GPR can accommodate nonlinear relationships while providing posterior predictive distributions that quantify uncertainty in estimated target values. Its probabilistic formulation also incorporates an observation-noise term during model fitting. Furthermore, GPR offers a practical modelling approach for limited experimental datasets \cite{xu_small_2023}, which is particularly relevant to process-engineering applications such as biomanufacturing, where high-quality data may be scarce \cite{peng_learning_2026}. Here, GPR provides a consistent basis for demonstrating how our automated feature-region selection pipeline can be integrated into soft-sensor development. A systematic comparison across regression models lies beyond the scope of this study, as the focus is on the automated identification and selection of informative contiguous regions in spectral-like measurements to support soft-sensor development in the process industries.

The proposed methodology is demonstrated using an electrochemical sensing platform for glucose concentration measurements. Using cyclic voltammetric measurements of glucose standards, the framework is evaluated within progressively broader concentration windows of 0--75, 0--150, 0--250, and 0--350~g/L. Glucose provides a broadly relevant case study for bioprocess monitoring, as it is widely employed as a carbon source and feedstock in bioprocesses. For example, in biofuel production, high- and very-high-gravity ethanol fermentations have been investigated using initial glucose concentrations of 150--350~g/L \cite{zhang_adaptive_2019,feng_development_2020}, whereas in biochemical production, batch fermentations for the production of succinate or butyric acid have been conducted at initial glucose concentrations ranging from 20 to 250~g/L \cite{qureshi_butyric_2022,chaleewong_kinetic_2022}. Reliable online estimation of glucose concentration could ultimately support process monitoring and closed-loop optimisation of feeding in bioprocesses.

The remainder of this paper is structured as follows. Section~\ref{sec:problem_definition} formulates the generalised soft-sensor development problem for spectral-like measurements. Section~\ref{sec:automated_framework} presents the automated feature-region selection framework, including candidate interval generation, model evaluation, and parsimonious final model selection, together with the GPR approach used in this study. Section~\ref{sec:glucose_case} describes the electrochemical glucose-sensing platform, generation of the experimental dataset, and application of the framework across the investigated nominal concentration ranges, including comparison with a GPR benchmark using the full measurement signal.

\section{Problem definition}
\label{sec:problem_definition}

Consider supervised data represented by $\mathcal{D}=(\mathbf{X},\mathbf{Y})$, where $\mathbf{X}\in\mathbb{R}^{n_d\times n_x}$ and $\mathbf{Y}\in\mathbb{R}^{n_d\times n_y}$. Here, $n_d$ denotes the number of experimental observations, $n_x$ the number of input features, and $n_y$ the number of target variables. Each row $\mathbf{x}_i\in\mathbb{R}^{n_x}$ of $\mathbf{X}$ represents a high-dimensional spectral-like measurement, while the corresponding row $\mathbf{y}_i\in\mathbb{R}^{n_y}$ of $\mathbf{Y}$ contains the target variables to be inferred.

A defining characteristic of a spectral-like measurement $\mathbf{x}_i$ is that its input features are ordered along an underlying physical measurement axis. We refer to the corresponding positions along this axis as measurement channels, collected in the
vector:
\begin{equation}
\boldsymbol{\xi}
=
\left[
\xi_1,\xi_2,\ldots,\xi_{n_x}
\right].
\end{equation}
The dimension of $\boldsymbol{\xi}$ therefore matches that of $\mathbf{x}_i$, with each feature $x_{i,j}$ of $\mathbf{x}_i$ associated with a measurement channel $\xi_j$. Depending on the sensing technology, $\boldsymbol{\xi}$ may correspond, for example, to wavenumber in spectroscopic measurements or applied potential in voltammetric measurements.

Using the complete set of input features, a general regression model $\mathcal{M}_R$ for the data in $\mathcal{D}$ can be expressed as:
\begin{equation}
\widehat{\mathbf{Y}}
=
\mathcal{M}_R
\left(
\mathbf{X}
\right),
\end{equation}
where $\widehat{\mathbf{Y}}$ denotes the predicted target variables.

As discussed in the Introduction, the complete feature matrix for high-dimensional spectral-like measurements may contain redundant, weakly informative, or noisy measurement regions. The objective considered in this work is therefore to construct more parsimonious soft sensors from automatically selected contiguous intervals of the original ordered feature space.

Let $\mathcal{I}_k$ denote the $k$th selected contiguous interval, defined as a contiguous set of ordered feature indices:
\begin{equation}
\mathcal{I}_k
=
\left\{
a_k,
a_k+1,
\ldots,
b_k
\right\},
\qquad
k=1,\ldots,n_I,
\end{equation}
where $a_k$ and $b_k$ denote the lower and upper indices, respectively, and $n_I$ denotes the number of final selected intervals. The collection of selected feature regions is then defined as:
\begin{equation}
\mathcal{E}
=
\left\{
\mathcal{I}_1,
\mathcal{I}_2,
\ldots,
\mathcal{I}_{n_I}
\right\},
\qquad
\mathcal{I}_k\cap\mathcal{I}_{\ell}
=
\varnothing,
\qquad
k\neq\ell,
\end{equation}
with the selected intervals constrained to be non-overlapping. The reduced feature matrix $\mathbf{X}_{\mathrm{sel}}$ containing only the features belonging to
the intervals in $\mathcal{E}$ is  $\mathbf{X}_{\mathrm{sel}}
\in
\mathbb{R}^{n_d\times n_{x,\mathrm{sel}}}$.

\textit{Remark on notation}. Let us introduce a generic selected-feature vector $\mathbf{x}_{\mathrm{sel}}\in
\mathbb{R}^{n_{x,\mathrm{sel}}}$ associated with a specific target variable. In the following, we focus on single-output regressors such that $n_y=1$; hence, $\mathbf{Y}:=\mathbf{y}\in\mathbb{R}^{n_d}$. Furthermore, let
$\widehat{y}$ denote the corresponding prediction of the target variable obtained from $\mathbf{x}_{\mathrm{sel}}$. The generalised regression model for a single target variable conditioned on the selected-feature vector $\mathbf{x}_{\mathrm{sel}}$ can then be written as:
\begin{equation}
\widehat{y}
=
\mathcal{M}_R
\left(
\mathbf{x}_{\mathrm{sel}}
\right).
\label{eq:regression_model}
\end{equation}

The question addressed in this work is then how to automatically identify informative contiguous feature regions to construct $\mathcal{E}$ for downstream soft-sensor development. The aim is for the resulting soft-sensor model to improve or at least retain the predictive performance obtained using the complete feature space when dealing with high-dimensional ordered spectral-like measurements. 

\section{Automated workflow for soft-sensor development}
\label{sec:automated_framework}

For the methodology, the complete dataset is partitioned into training, validation, and test sets using an approximately 60\%, 20\%, and 20\% split, respectively. The partitioning is performed based on unique target-variable levels, such that all replicate measurements corresponding to a given level are assigned exclusively to a single subset, thereby avoiding replicate leakage across the training,
validation, and test sets. 

\subsection{Point-wise scoring}
\label{subsec:point_wise_scoring}

Using the \textit{training} set, the first step of the automated workflow quantifies the information value associated with each individual measurement channel $\xi_j$, $j=1,\ldots,n_x$. For a scalar target variable, let $y_g$, $g=1,\ldots,n_G$, denote the $n_G$ distinct target-variable levels represented in the training set; in the case study, these represent glucose concentration levels. Then, let $x_{g,r,j}$ denote the measured value at channel $\xi_j$ for replicate $r$ at target-variable level $y_g$. Considering $n_R$ replicate measurements at each target-variable level, the mean response
at channel $\xi_j$ is calculated as:
\begin{equation}
\bar{x}_{g,j}
=
\frac{1}{n_R}
\sum_{r=1}^{n_R}
x_{g,r,j}.
\end{equation}

The predictive relevance of each measurement channel is quantified through the squared Spearman's rank correlation coefficient between the target-variable levels and their corresponding mean responses:
\begin{equation}
\rho_{s,j}^{2}
=
\operatorname{corr}_{\mathrm{S}}^{2}
\left(
\left\{y_g\right\}_{g=1}^{n_G},
\left\{\bar{x}_{g,j}\right\}_{g=1}^{n_G}
\right),
\end{equation}
where $\operatorname{corr}_{\mathrm{S}}$ denotes Spearman's rank correlation. Spearman's rank correlation was used because of its ability to quantify whether two variables are monotonically related regardless of whether the relationship is linear or nonlinear.
Squaring the coefficient ensures that both positive and negative monotonic relationships contribute equally to the measure of predictive relevance. A limitation is that it does not capture general non-monotonic relationships.

A second measurement-quality metric, $RV_j$, is defined by comparing the variation in the mean response across target-variable levels with the average variation observed among experimental replicates for the corresponding channel $\xi_j$:
\begin{equation}
RV_j
=
\frac{
\operatorname{Var}
\left(
\left\{\bar{x}_{g,j}\right\}_{g=1}^{n_G}
\right)
}{
\displaystyle
\frac{1}{n_G}
\sum_{g=1}^{n_G}
\operatorname{Var}
\left(
\left\{x_{g,r,j}\right\}_{r=1}^{n_R}
\right)
+
\varepsilon
},
\end{equation}
where $\operatorname{Var}$ is the sample variance and $\varepsilon$ is a small positive constant introduced to avoid the possibility of division by zero. A large $RV_j$ therefore indicates that the response at measurement channel $\xi_j$ varies substantially across target-variable levels relative to the average variability observed between replicate measurements. This serves as a signal-to-noise indicator.

The predictive-relevance and measurement-quality metrics are subsequently combined into the point-wise composite score:
\begin{equation}
S_j
=
\rho_{s,j}^{2}
\log
\left(
1+RV_j
\right).
\label{eq:inf_channel}
\end{equation}
Here, $\log$ denotes the natural logarithm. The resulting sequence, $\mathbf{S}
=
\left[
S_1,S_2,\ldots,S_{n_x}
\right]$, provides a fingerprint revealing a latent information pattern along the measurement axis $\boldsymbol{\xi}$. Measurement channels presenting both a strong monotonic correlation with the target variable and substantial variation across target-variable levels relative to replicate variability consequently receive higher scores. This point-wise information profile provides the basis for the subsequent identification and scoring of contiguous candidate feature regions.

\subsection{Candidate interval generation from the information fingerprint}
\label{subsec:candidate_int_gen}

From the point-wise information-score profile as a function of measurement channel, generated from the \textit{training} set, dominant peaks are automatically identified using peak-prominence and minimum peak-distance criteria. Only peaks whose prominence reaches a prescribed fraction of the maximum point-wise information score are retained.

Around each retained peak, candidate measurement-channel regions are generated using the relative-height concept, whereby peak widths are evaluated at different fractions of the peak prominence to define intervals of varying extent around the peak. Each candidate region is denoted as:
\begin{equation}
\mathcal{C}_{p,m}
=
\left\{
a_{p,m},
a_{p,m}+1,
\ldots,
b_{p,m}
\right\},
\end{equation}
where $p$ denotes the detected peak, $m$ denotes the relative-height value used to define the candidate region, and $a_{p,m}$ and $b_{p,m}$ are its lower and upper ordered feature indices, respectively. In the case study, peak detection and width estimation are performed using
\texttt{scipy.signal.find\_peaks} and \texttt{scipy.signal.peak\_widths} \cite{2020SciPy-NMeth}. In the case study, a minimum peak-distance criterion of 5 measurement channels is employed, and the prominence threshold is set to 0.15 times the maximum point-wise information score. Furthermore, the relative-height values $\{0.25,0.40,0.50,0.65,0.75,0.85\}$ are evaluated.

To favour local feature regions over isolated point-wise spikes, only intervals containing a prescribed minimum number of contiguous measurement channels are retained; in the case study, a minimum of 3 channels is required. This ensures that retained intervals represent local measurement-channel regions rather than isolated point-wise responses. Although isolated channels may themselves contain predictive information, here we deliberately restrict the automated selection to contiguous feature regions to preserve the structured, ordered nature of spectral-like signals.

For each automatically defined interval, we compute an interval-wise score as the average point-wise score: 
\begin{equation}
\bar{S}_{p,m}
=
\frac{1}{\left|\mathcal{C}_{p,m}\right|}
\sum_{j\in\mathcal{C}_{p,m}}
S_j.
\end{equation}
This metric measures the average information richness of the candidate feature interval, as it inherently accounts for the predictive-relevance and measurement-quality contributions incorporated in $S_j$ (cf. Eq.~\eqref{eq:inf_channel}).

For parsimony, among candidate intervals originating from the same detected peak but defined using different relative-height values, only the interval with the highest average information score is retained for downstream processing, subject to the prescribed minimum interval size. In the event of a tie, the interval containing fewer measurement channels is retained. A consistency check then verifies that the retained intervals do not overlap within each signal branch.

Let $\mathcal{A}$ denote the index set of candidate intervals retained
after the interval-wise filtering step. The retained interval collection is:
\begin{equation}
\left\{
\mathcal{C}_{p,m}
\;:\;
(p,m)\in\mathcal{A}
\right\}.
\end{equation}

Combinations of candidate measurement-channel regions are then generated by considering individual retained intervals and unions of retained intervals, up to a user-defined maximum cardinality $Q\leq|\mathcal{A}|$:
\begin{equation}
\mathfrak{C}_q
=
\left\{
\bigcup_{(p,m)\in\mathcal{J}}
\mathcal{C}_{p,m}
\;:\;
\mathcal{J}\subseteq\mathcal{A},
\quad
|\mathcal{J}|=q
\right\},
\qquad
q=1,\ldots,Q.
\end{equation}
The complete set of candidate feature-region combinations is
\begin{equation}
\mathfrak{C}
=
\bigcup_{q=1}^{Q}
\mathfrak{C}_q.
\end{equation}
In the case study, $Q=|\mathcal{A}|$, such that all combinations of the retained intervals indexed by $\mathcal{A}$ are evaluated.

\subsection{Ranking of soft-sensor candidates and final model selection}

Let $\mathcal{K}=\{1,\ldots,n_C\}$ index the $n_C=|\mathfrak{C}|$ candidate feature-region combinations. Each candidate $c\in\mathcal{K}$ is used to construct a selected-feature matrix
$\mathbf{X}_{\mathrm{sel},c}$, and a regression model $\mathcal{M}_{R,c}$ is fitted using the \textit{training} set (cf. Eq.~\eqref{eq:regression_model}). Its predictive performance is
then evaluated using the held-out \textit{validation} set. Candidate models are ranked according to their validation root mean squared error (RMSE).

The minimum validation RMSE among the candidate models is first identified:
\begin{equation}
\mathrm{RMSE}_{\mathrm{val}}^{\min}
=
\min_{c\in\mathcal{K}}
\mathrm{RMSE}_{\mathrm{val},c}.
\end{equation}

To favour parsimony in model selection, candidates presenting a validation RMSE within a prescribed relative tolerance $\tau$ of this minimum are considered to provide comparable predictive performance:
\begin{equation}
\mathcal{K}_{\mathrm{tol}}
=
\left\{
c\in\mathcal{K}
\;:\;
\mathrm{RMSE}_{\mathrm{val},c}
\leq
(1+\tau)
\mathrm{RMSE}_{\mathrm{val}}^{\min}
\right\}.
\end{equation}

From this subset, the candidate containing the smallest number of input features, i.e., measurement channels, is selected:
\begin{equation}
c^{*}
\in
\underset{c\in\mathcal{K}_{\mathrm{tol}}}{\arg\min}
\;
n_{x,\mathrm{sel},c}.
\end{equation}
This prevents the selection of an unnecessarily complex model when only a marginal improvement in validation performance is obtained. If multiple candidates contain the same minimum number of input features, the candidate with the lowest validation RMSE is selected. The selected candidate $c^{*}$ determines the final collection of feature intervals $\mathcal{E}$. In the case study, a relative tolerance of $\tau=0.025$ is employed.

Following model selection, the regression model associated with $c^{*}$ is refitted using the combined training and validation sets and denoted by $\mathcal{M}_{R}^{*}$. The held-out \textit{test} set remains unused during feature and model selection and is reserved for
the final evaluation of predictive performance on unseen target-variable levels.

\subsection{Gaussian-process-based soft sensors}

Let $d=n_{x,\mathrm{sel}}$ denote the selected-feature dimension.
Single-output GPR \cite{rasmussen_gaussian_2006} models an unknown
function $f:\mathbb{R}^{d}\rightarrow\mathbb{R}$. The measured target value associated with a selected-feature vector is
assumed to be a noisy observation of this function:
\begin{equation}
y
=
f\left(\mathbf{x}_{\mathrm{sel}}\right)
+
\eta,
\qquad
\eta
\sim
\mathcal{N}\left(0,\sigma_n^2\right),
\end{equation}
where $\sigma_n^2$ denotes the observation-noise variance. GPR defines a probability distribution over the function $f$:
\begin{equation}
f
\sim
\mathcal{GP}
\left(
\mathfrak{m}(\mathbf{x}_{\mathrm{sel}}),
k(\mathbf{x}_{\mathrm{sel}},\mathbf{x}_{\mathrm{sel}}')
\right),
\end{equation}
where $\mathfrak{m}(\mathbf{x}_{\mathrm{sel}})$ is the prior mean
function and
$k(\mathbf{x}_{\mathrm{sel}},\mathbf{x}_{\mathrm{sel}}')$ is the
covariance or kernel function evaluated between two arbitrary
selected-feature vectors $\mathbf{x}_{\mathrm{sel}}$ and
$\mathbf{x}_{\mathrm{sel}}'$. The prior mean and covariance functions
jointly define the GP prior.

For the $n$ observations used to fit a given GPR model, let
$\mathbf{x}_{\mathrm{sel},1},\ldots,
\mathbf{x}_{\mathrm{sel},n}$ denote the corresponding selected-feature
vectors. Let us introduce the kernel matrix, which contains the pairwise
kernel evaluations among the $n$ input vectors used for model fitting:
\begin{equation}
\mathbf{K}
=
\begin{bmatrix}
k(\mathbf{x}_{\mathrm{sel},1},\mathbf{x}_{\mathrm{sel},1})
& \cdots &
k(\mathbf{x}_{\mathrm{sel},1},\mathbf{x}_{\mathrm{sel},n})
\\
\vdots & \ddots & \vdots
\\
k(\mathbf{x}_{\mathrm{sel},n},\mathbf{x}_{\mathrm{sel},1})
& \cdots &
k(\mathbf{x}_{\mathrm{sel},n},\mathbf{x}_{\mathrm{sel},n})
\end{bmatrix}
\in\mathbb{R}^{n\times n}.
\end{equation}

In the case study, a radial basis function (RBF) kernel is used:
\begin{equation}
k
\left(
\mathbf{x}_{\mathrm{sel}},
\mathbf{x}_{\mathrm{sel}}'
\right)
=
\sigma_f^2
\exp
\left[
-\frac{1}{2}
\left(
\mathbf{x}_{\mathrm{sel}}
-
\mathbf{x}_{\mathrm{sel}}'
\right)^{\top}
\mathbf{M}
\left(
\mathbf{x}_{\mathrm{sel}}
-
\mathbf{x}_{\mathrm{sel}}'
\right)
\right],
\label{eq:kernel_function}
\end{equation}
with $\mathbf{M}
=
\operatorname{diag}
\left(
\ell_1^{-2},
\ldots,
\ell_d^{-2}
\right)$. Here, $\sigma_f^2$ denotes the signal variance and
$\boldsymbol{\ell}
=
[\ell_1,\ldots,\ell_d]^{\top}$
denotes the length-scale vector. Note that a separate length scale is fitted for each input feature, corresponding to automatic relevance determination
(ARD).

A linear prior mean function is employed:
\begin{equation}
\mathfrak{m}
\left(
\mathbf{x}_{\mathrm{sel}}
\right)
=
\mathbf{x}_{\mathrm{sel}}^{\top}\mathbf{w}
+
b,
\end{equation}
where $\mathbf{w}\in\mathbb{R}^{d}$ is the weight vector and
$b\in\mathbb{R}$ is a constant bias. Let
$\mathbf{y}=[y_1,\ldots,y_n]^{\top}$ collect the $n$ target observations and let
$\mathbf{m}\in\mathbb{R}^{n}$ collect the prior-mean evaluations, such
that:
\begin{equation}
\mathbf{m}
=
\left[
\mathfrak{m}(\mathbf{x}_{\mathrm{sel},1}),
\ldots,
\mathfrak{m}(\mathbf{x}_{\mathrm{sel},n})
\right]^{\top}.
\end{equation}

The Gaussian log marginal likelihood is:
\begin{equation}
\begin{split}
\log p
\left(
\mathbf{y}
\mid
\mathbf{X}_{\mathrm{sel}},
\boldsymbol{\theta}
\right)
={}&
-\frac{1}{2}
(\mathbf{y}-\mathbf{m})^{\top}
\left(
\mathbf{K}+\sigma_n^2\mathbf{I}_n
\right)^{-1}
(\mathbf{y}-\mathbf{m})
\\
&-
\frac{1}{2}
\log
\left|
\mathbf{K}+\sigma_n^2\mathbf{I}_n
\right|
-
\frac{n}{2}\log(2\pi),
\end{split}
\end{equation}
where $\mathbf{I}_n$ is the $n$-dimensional identity matrix. 

The trainable model parameters are collected in:
\begin{equation}
\boldsymbol{\theta}
=
\left[
\sigma_n^2,
\sigma_f^2,
\boldsymbol{\ell}^{\top},
\mathbf{w}^{\top},
b
\right]^{\top}.
\end{equation}
These parameters are fitted jointly by maximising the Gaussian log marginal likelihood.

For a query selected-feature vector
$\mathbf{x}_{\mathrm{sel}}^{*}$, let us define the covariance vector:
\begin{equation}
\mathbf{k}_{*}
=
\left[
k(\mathbf{x}_{\mathrm{sel},1},\mathbf{x}_{\mathrm{sel}}^{*}),
\ldots,
k(\mathbf{x}_{\mathrm{sel},n},\mathbf{x}_{\mathrm{sel}}^{*})
\right]^{\top}.
\end{equation}

The posterior mean and variance of the function $f$ are then given,
respectively, by:
\begin{equation}
\mu_{*}
\left(
\mathbf{x}_{\mathrm{sel}}^{*}
\right)
=
\mathfrak{m}
\left(
\mathbf{x}_{\mathrm{sel}}^{*}
\right)
+
\mathbf{k}_{*}^{\top}
\left(
\mathbf{K}+\sigma_n^2\mathbf{I}_n
\right)^{-1}
(\mathbf{y}-\mathbf{m}),
\end{equation}
and
\begin{equation}
s_{*}^{2}
\left(
\mathbf{x}_{\mathrm{sel}}^{*}
\right)
=
k
\left(
\mathbf{x}_{\mathrm{sel}}^{*},
\mathbf{x}_{\mathrm{sel}}^{*}
\right)
-
\mathbf{k}_{*}^{\top}
\left(
\mathbf{K}+\sigma_n^2\mathbf{I}_n
\right)^{-1}
\mathbf{k}_{*}.
\end{equation}
The posterior variance $s_{*}^{2}$ quantifies epistemic uncertainty in
the function prediction, conditional on the fitted GP parameters
and the observations used for training.

In the implementation, GPR models were developed using
\texttt{SingleTaskGP} from BoTorch \cite{balandat2020botorch}, with
GPyTorch modules \cite{gardner2018gpytorch}. Model parameters are fitted using
\texttt{ExactMarginalLogLikelihood} and
\texttt{fit\_gpytorch\_mll}. The default \texttt{SingleTaskGP} outcome transformation standardises
the target observations during model fitting, and posterior quantities
are subsequently returned on the original target scale. No
transformation is applied to the input features.
Posterior predictions are obtained without observation noise, such that the reported posterior variance only quantifies uncertainty in the function $f$.

\section{Case study: glucose soft sensing from voltammetric measurements}
\label{sec:glucose_case}

\begin{figure*}[htb]
    \centering
    \includegraphics[width=0.8\linewidth]{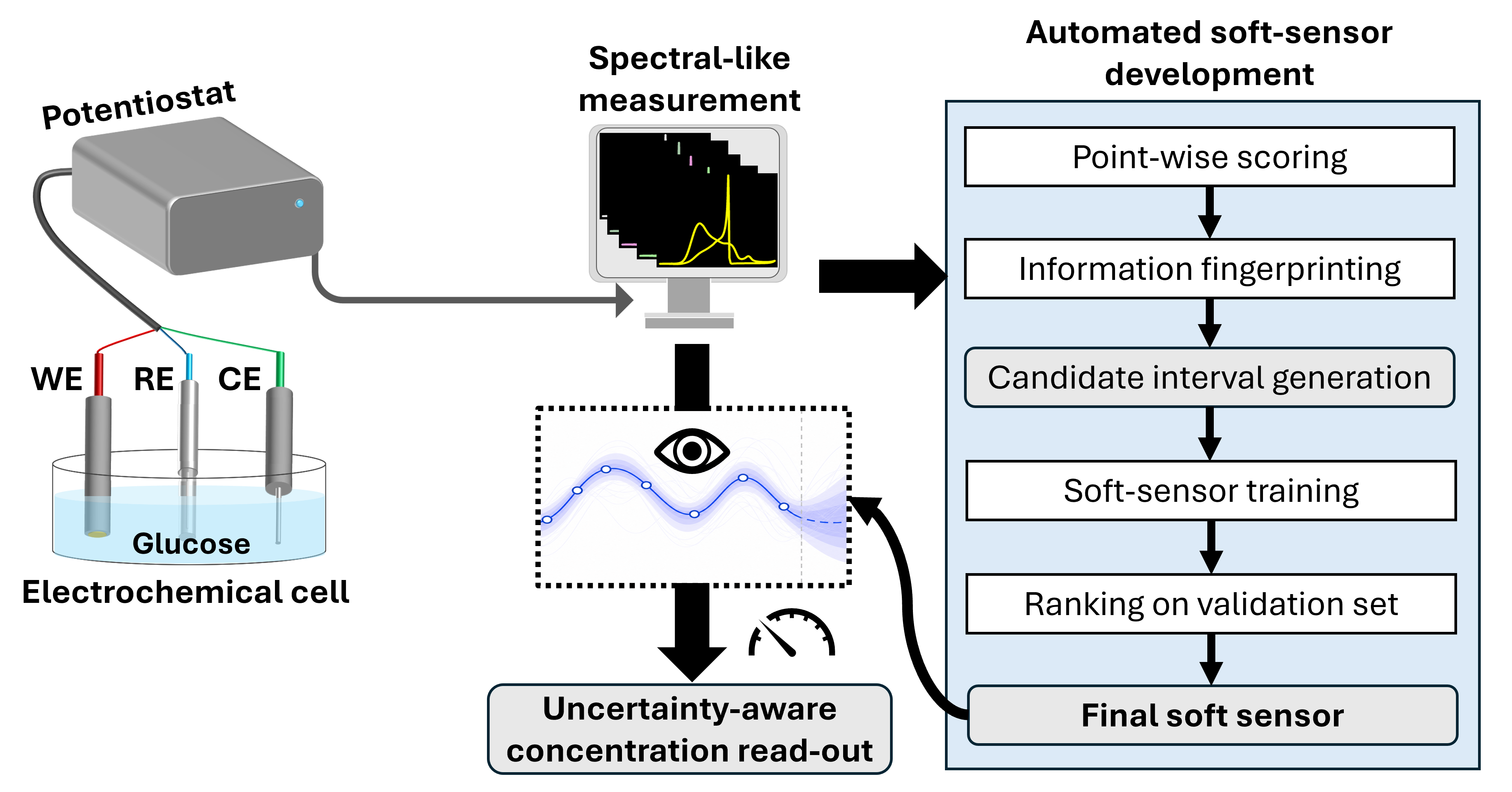}
    \caption{Overview of the glucose soft-sensing case study, including the three-electrode experimental setup used to acquire spectral-like voltammetric measurements and the automated soft-sensor development pipeline. WE: working electrode; RE: reference electrode; CE: counter
    electrode.}
    \label{fig:overview_case_study}
\end{figure*}

To demonstrate the automated soft-sensor development framework for spectral-like measurements, we apply it to voltammetric measurements of glucose. Fig.~\ref{fig:overview_case_study} provides an overview of the experimental setup and the application of the framework considered in this case study.

\subsection{Voltammetric measurements of glucose using an Au electrode}

Cyclic voltammograms were recorded using an Au working electrode in
0.1 M phosphate buffer solution (PBS, pH = 13) at the nominal glucose concentrations listed in Table~\ref{tab:data_splits}. Details on the mechanism underlying the electrochemical response of glucose on gold surfaces can be found in \cite{kirk_electrochemical_1980,pasta_mechanism_2010}.

\begin{table*}[htb!]
\centering
\caption{Glucose concentration levels assigned to the training, validation, and test sets for each investigated nominal sensing range. All replicate measurements corresponding to a given concentration level were assigned
to the same subset.}
\label{tab:data_splits}

\footnotesize
\setlength{\tabcolsep}{5pt}

\begin{tabularx}{\textwidth}{
    @{}
    >{\raggedright\arraybackslash}p{0.10\textwidth}
    >{\raggedright\arraybackslash}X
    >{\raggedright\arraybackslash}X
    >{\raggedright\arraybackslash}X
    @{}
}
\toprule
\textbf{Range [g/L]} &
\textbf{Training concentrations [g/L]} &
\textbf{Validation concentrations [g/L]} &
\textbf{Test concentrations [g/L]} \\
\midrule

0--75 &
0,  2,  7,  9, 11, 13, 17, 19, 20, 22, 24, 28, 30, 32, 35, 37, 50, 51,
 68 &
6, 26, 33, 39, 41, 42 &
4, 15, 43, 44, 46, 48, 59 \\[1mm]

0--150 &
0,   2,   4,   7,   9,  13,  15,  17,  19,  20,  22,  24,  26,  30,
  33,  41,  42,  68,  77, 103, 120, 129, 137, 146 &
37,  43,  44,  48,  50,  51,  94, 111 &
6, 11, 28, 32, 35, 39, 46, 59, 86 \\[1mm]

0--250 &
0,   2,   4,   7,  13,  15,  17,  19,  22,  24,  26,  28,  35,  41,
  50,  51,  68,  86,  94, 103, 120, 129, 137, 146, 164, 181, 189, 198,
 207, 224, 250 &
6,  11,  20,  30,  42,  43,  59, 111, 172, 215, 241 &
9,  32,  33,  37,  39,  44,  46,  48,  77, 154, 232 \\[1mm]

0--350 &
0, 2, 4, 7, 9, 13, 15, 17, 19, 22, 24, 26, 28, 33, 41, 43, 51,
68, 86, 94, 103, 120, 129, 137, 146, 164, 172, 181, 189, 198,
207, 224, 259, 283, 296 &
6, 11, 30, 44, 48, 50, 59, 111, 154, 215, 250, 322 &
20, 32, 35, 37, 39, 42, 46, 77, 232, 241, 267, 348 \\

\bottomrule
\end{tabularx}
\end{table*}

All electrochemical measurements were conducted at room temperature using a conventional 20-mL three-electrode electrochemical cell. The configuration consisted of a platinum counter electrode, an Ag/AgCl reference electrode, and a 2-mm-diameter Au working electrode (CH Instruments, Austin, TX, USA). The Au working electrode was polished with 0.05-\textmu m alumina slurry, thoroughly rinsed with pure water, and dried at room temperature before use. Electrochemical measurements were performed using an eight-channel potentiostat (EmStat 4 LR; PalmSens, The Netherlands). Cyclic voltammograms were recorded at a scan rate of 50~mV~s$^{-1}$ over a potential range from -1.0 to 1.0~V, with a potential step of 0.01~V. Five replicate measurements, each comprising two voltammetric cycles, were recorded at each glucose concentration. For dataset construction, the second cycle was retained at 0 g/L and the first cycle at all other concentrations.

In the experimental setup, all reagents were of high purity, analytical grade, or equivalent. Phosphoric acid and sodium hydroxide (Sigma-Aldrich) were used to prepare the 0.1 M phosphate buffer solution (PBS). D-(+)-Glucose (dextrose) anhydrous (Sigma-Aldrich) was dissolved in 0.1 M PBS to prepare the analyte solutions used for electrochemical analysis.

\begin{figure*}[htb!]
\makebox[\textwidth][c]{%
\makebox[0.32\textwidth]{\textbf{A) Complete voltammetric cycle}}
\makebox[0.32\textwidth]{\textbf{B) Oxidation branch}}
\makebox[0.32\textwidth]{\textbf{C) Reduction branch}}
}
\\[1mm]
\centering
    \begin{subfigure}{0.32\textwidth}
      \captionsetup{justification=centering}
      \includegraphics[scale=0.43]{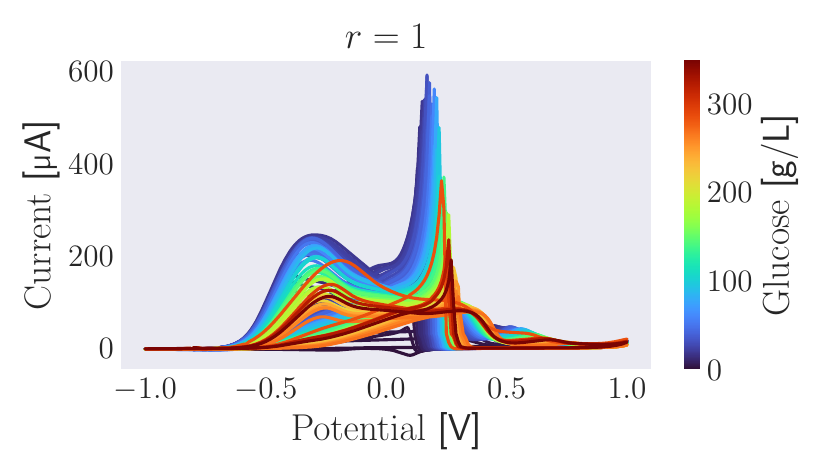}
    \end{subfigure}
    \begin{subfigure}{0.32\textwidth}
      \captionsetup{justification=centering}
      \includegraphics[scale=0.43]{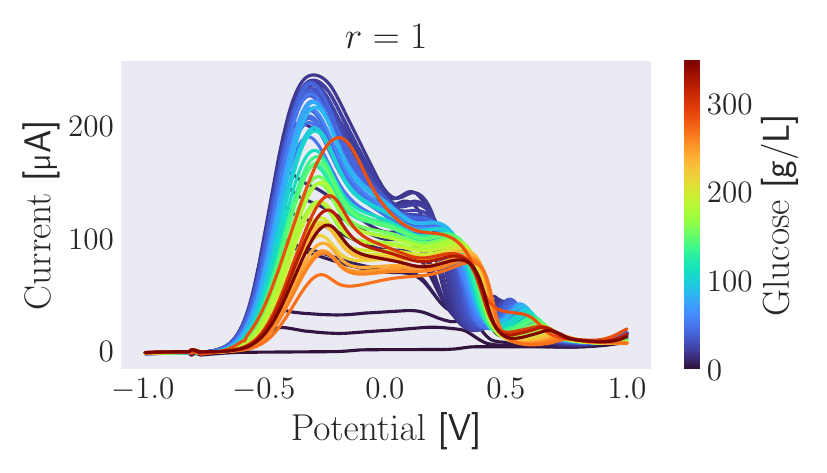}
    \end{subfigure}
    \begin{subfigure}{0.32\textwidth}
      \captionsetup{justification=centering}
      \includegraphics[scale=0.43]{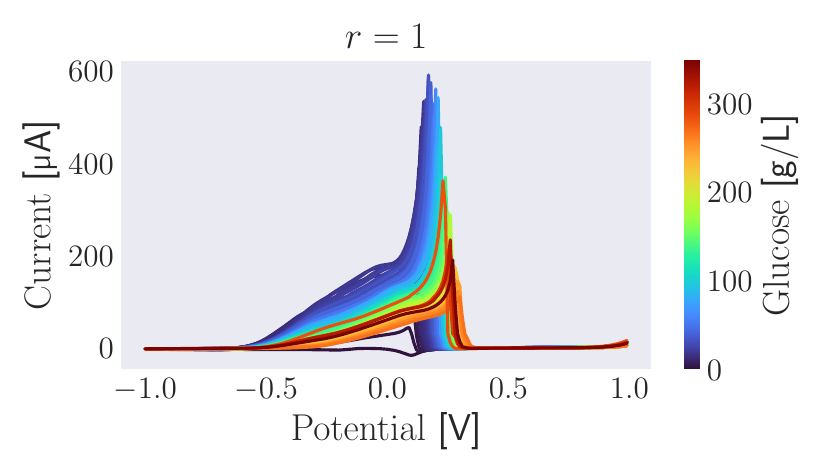}
    \end{subfigure}
    \\
        \begin{subfigure}{0.32\textwidth}
      \captionsetup{justification=centering}
      \includegraphics[scale=0.43]{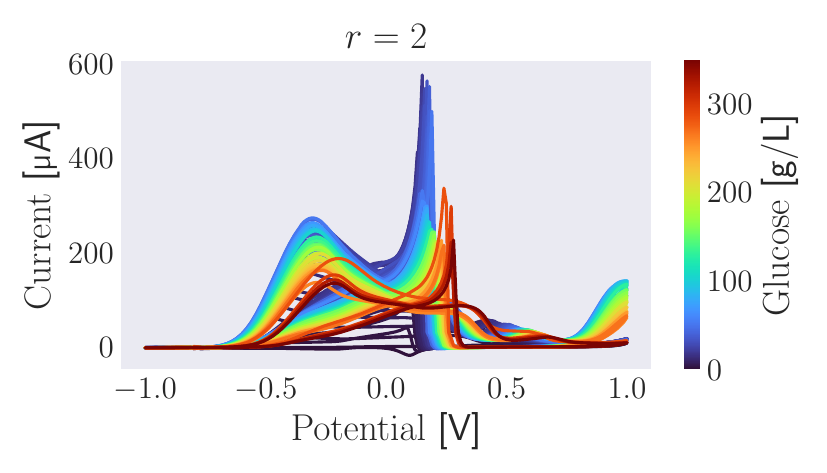}
    \end{subfigure}
    \begin{subfigure}{0.32\textwidth}
      \captionsetup{justification=centering}
      \includegraphics[scale=0.43]{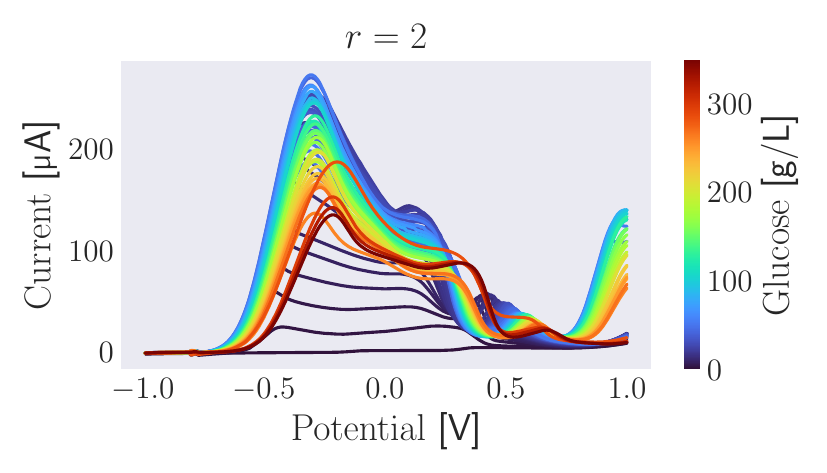}
    \end{subfigure}
    \begin{subfigure}{0.32\textwidth}
      \captionsetup{justification=centering}
      \includegraphics[scale=0.43]{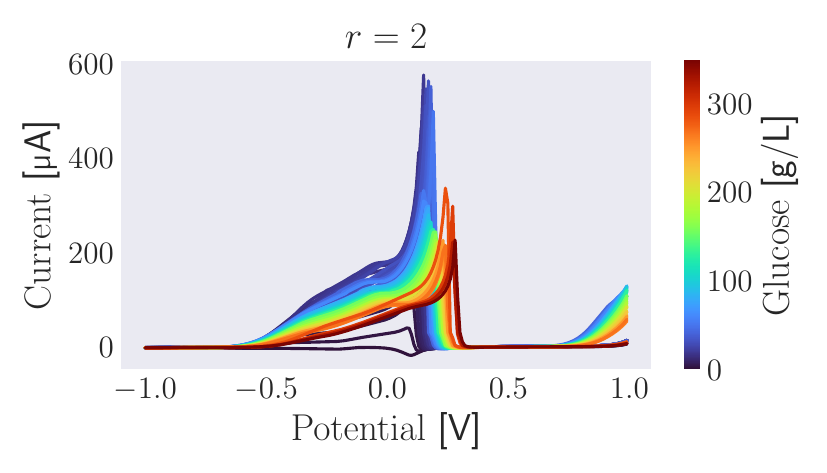}
    \end{subfigure}
    \\
    \begin{subfigure}{0.32\textwidth}
      \captionsetup{justification=centering}
      \includegraphics[scale=0.43]{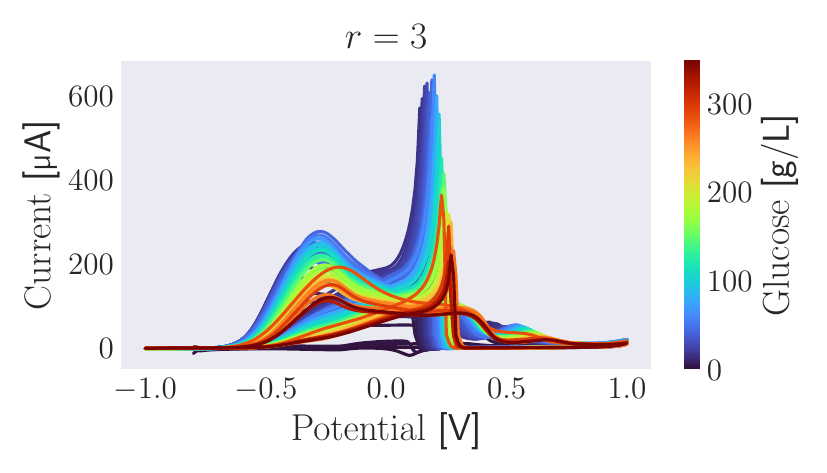}
    \end{subfigure}
    \begin{subfigure}{0.32\textwidth}
      \captionsetup{justification=centering}
      \includegraphics[scale=0.43]{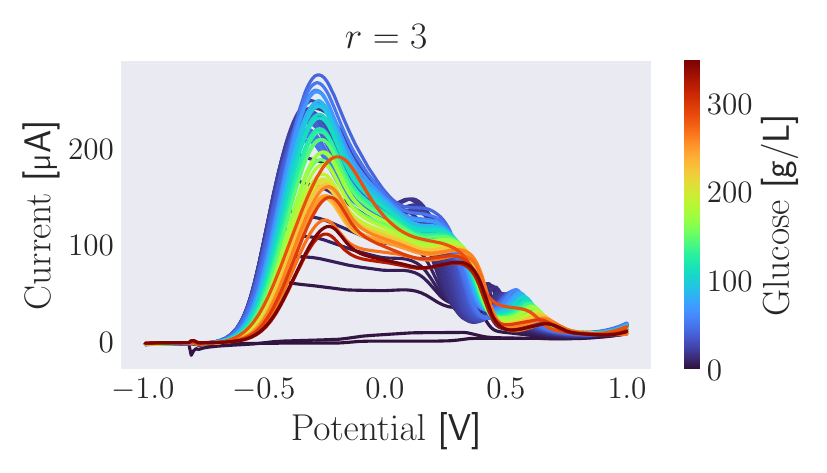}
    \end{subfigure}
    \begin{subfigure}{0.32\textwidth}
      \captionsetup{justification=centering}
      \includegraphics[scale=0.43]{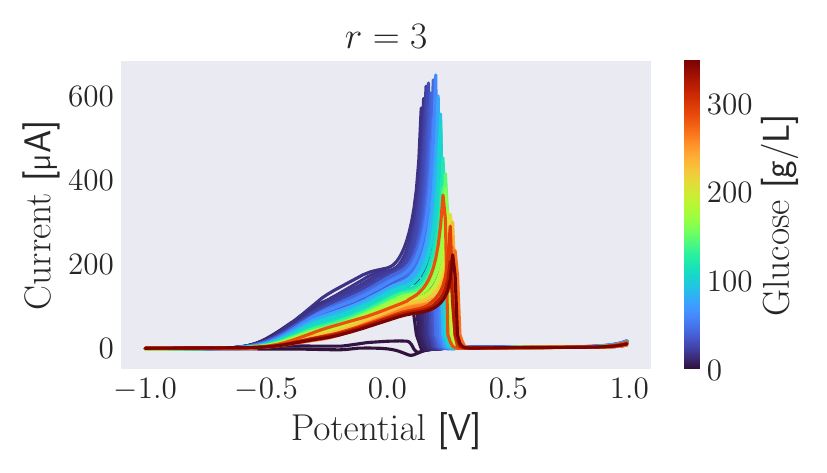}
    \end{subfigure}
    \\
    \begin{subfigure}{0.32\textwidth}
      \captionsetup{justification=centering}
      \includegraphics[scale=0.43]{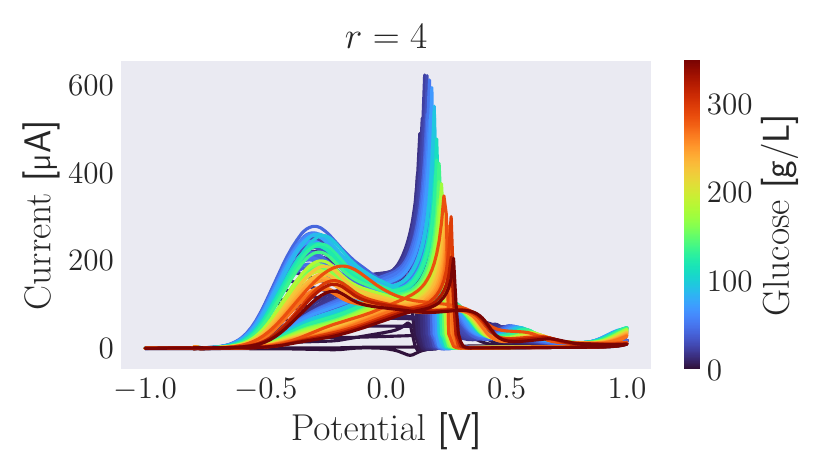}
    \end{subfigure}
    \begin{subfigure}{0.32\textwidth}
      \captionsetup{justification=centering}
      \includegraphics[scale=0.43]{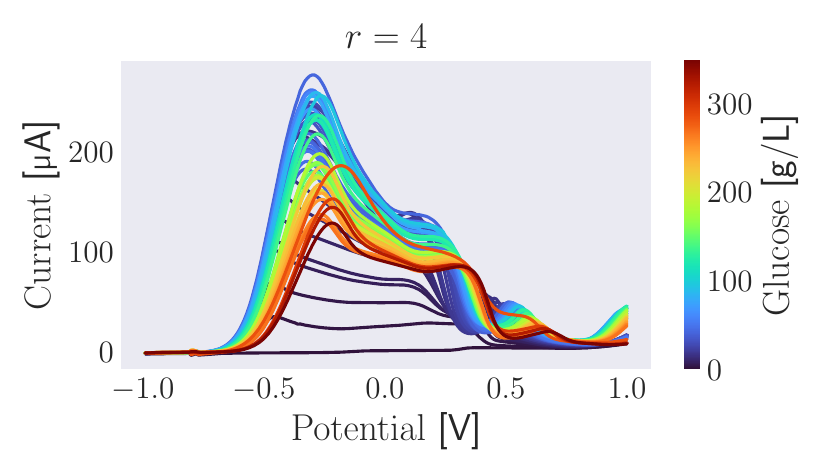}
    \end{subfigure}
    \begin{subfigure}{0.32\textwidth}
      \captionsetup{justification=centering}
      \includegraphics[scale=0.43]{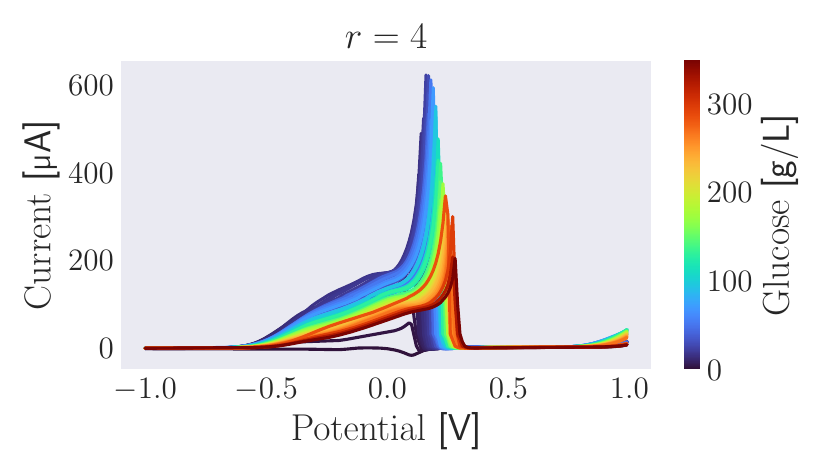}
    \end{subfigure}
        \\
    \begin{subfigure}{0.32\textwidth}
      \captionsetup{justification=centering}
      \includegraphics[scale=0.43]{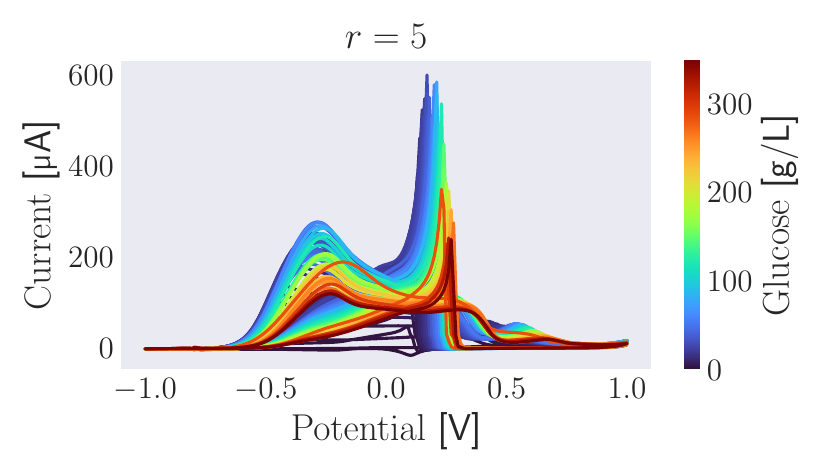}
    \end{subfigure}
    \begin{subfigure}{0.32\textwidth}
      \captionsetup{justification=centering}
      \includegraphics[scale=0.43]{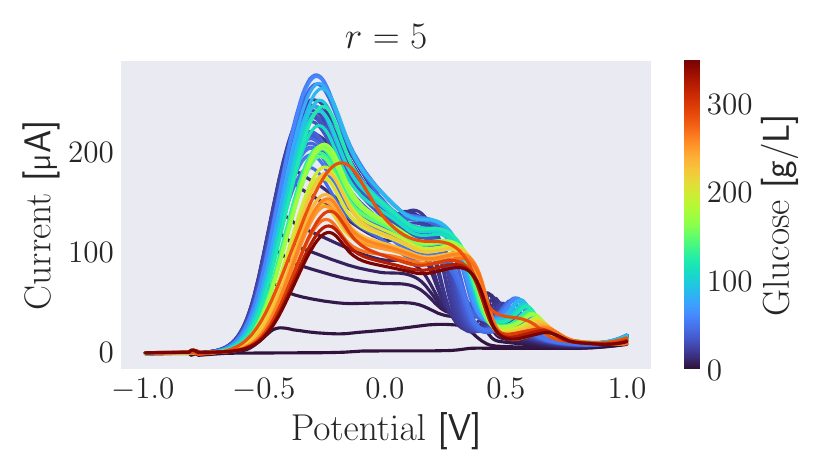}
    \end{subfigure}
    \begin{subfigure}{0.32\textwidth}
      \captionsetup{justification=centering}
      \includegraphics[scale=0.43]{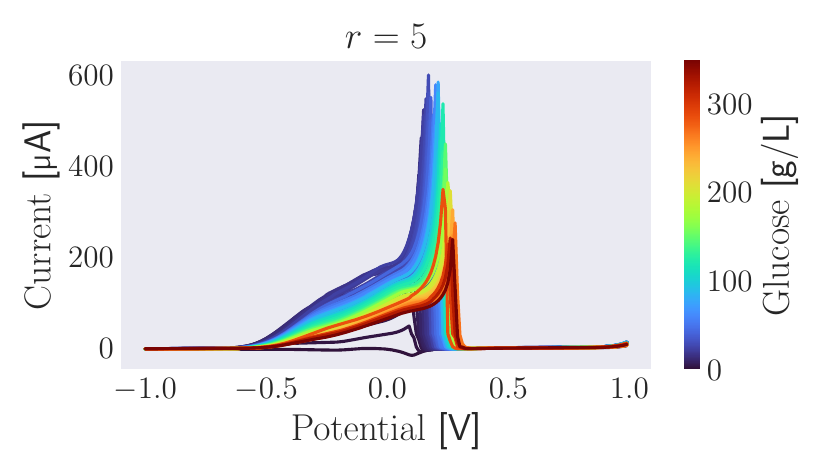}
    \end{subfigure}
    \\
\caption{Cyclic voltammetry measurements across the 0--350 g/L
nominal glucose concentration range for five experimental replicates ($r=1,\ldots,5$).
A) Complete cyclic voltammogram readouts.
B) Oxidation-channel readouts.
C) Reduction-channel readouts.
Line colour denotes glucose concentration.}
    \label{fig:features_GPS}
\end{figure*}

A cyclic voltammogram contains one branch acquired as the working-electrode potential is swept towards more positive values and another acquired as the potential is swept towards more negative values. These potential sweeps alter the driving force for electron transfer and are denoted here as the oxidation and reduction branches, respectively. Because the same applied potential can be sampled on both branches while producing different current responses, the corresponding measurements are treated as distinct channels and labelled accordingly. Both sets of channels are included in each spectral-like measurement $\mathbf{x}_i$; thus, each channel coordinate $\xi_j$ identifies a potential--branch pair. After branch separation, each processed voltammogram contained 200 oxidation and 200 reduction channels, giving 400 measurement channels in total. In Fig.~\ref{fig:features_GPS}, we show the full cyclic voltammogram readouts, as well as the separate oxidation- and reduction-channel readouts.

Notably, retaining both branches increases the input feature dimension relative to a representation containing only one current response per applied potential, providing a more challenging case study for evaluating the automated feature-selection framework for soft-sensor development. Although the voltammograms contain visually apparent peaks, it is not immediately evident which intervals or interval combinations will yield the best predictive performance, nor whether the largest current peaks are the most informative under the proposed scoring framework. 

In the following subsections, the automated feature-selection framework is applied across progressively broader nominal glucose concentration ranges. For reference, the glucose concentration levels assigned to the training, validation, and test subsets for each nominal range are listed in Table~\ref{tab:data_splits}.

\subsection{Automated soft-sensor development for the 0--75 g/L range}

\begin{figure*}[htb!]
\centering

% Column headings
\makebox[\textwidth][c]{%
\makebox[0.23\textwidth][c]{\textbf{A) 0--75 g/L}}%
\makebox[0.23\textwidth][c]{\textbf{B) 0--150 g/L}}%
\makebox[0.23\textwidth][c]{\textbf{C) 0--250 g/L}}%
\makebox[0.23\textwidth][c]{\textbf{D) 0--350 g/L}}%
}
\\[1mm]

% ---------------------------------------------------------------------
% OXIDATION: squared Spearman correlation
% ---------------------------------------------------------------------
\begin{subfigure}{0.23\textwidth}
    \centering
    \includegraphics[width=\linewidth]{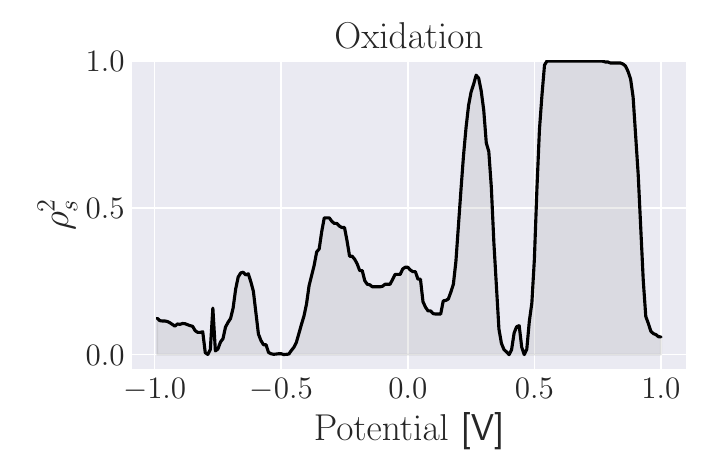}
\end{subfigure}
\begin{subfigure}{0.23\textwidth}
    \centering
    \includegraphics[width=\linewidth]{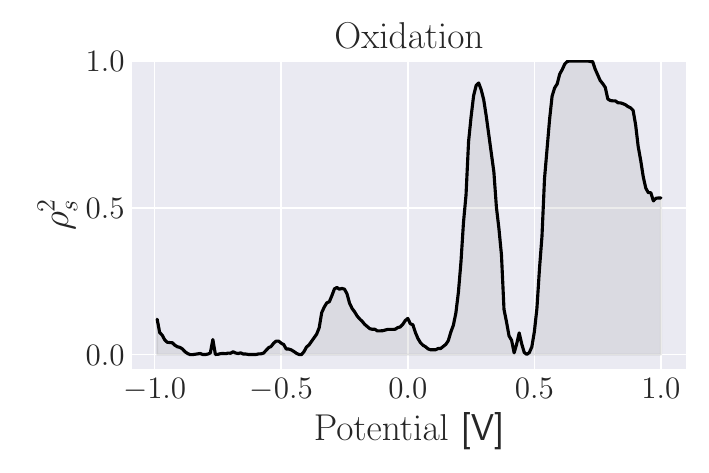}
\end{subfigure}
\begin{subfigure}{0.23\textwidth}
    \centering
    \includegraphics[width=\linewidth]{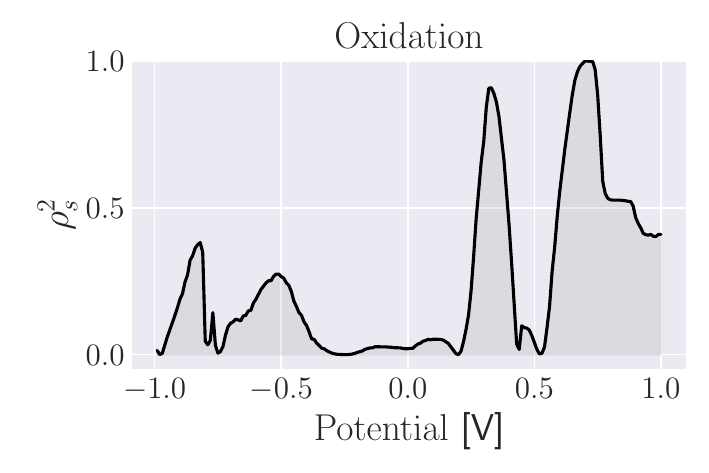}
\end{subfigure}
\begin{subfigure}{0.23\textwidth}
    \centering
    \includegraphics[width=\linewidth]{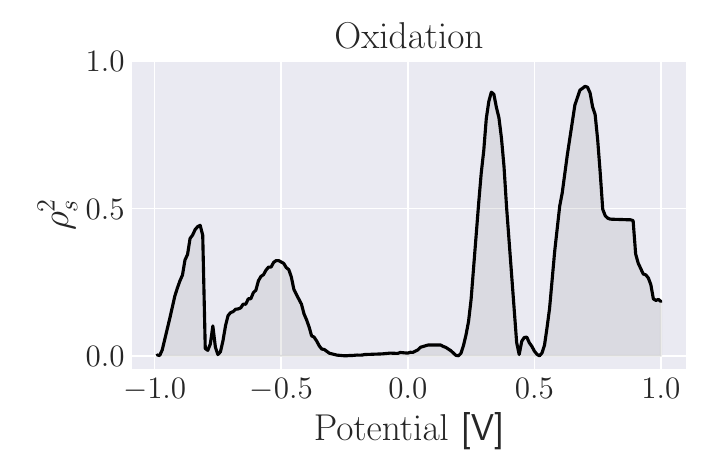}
\end{subfigure}
\\

% ---------------------------------------------------------------------
% OXIDATION: measurement-quality metric
% ---------------------------------------------------------------------
\begin{subfigure}{0.23\textwidth}
    \centering
    \includegraphics[width=\linewidth]{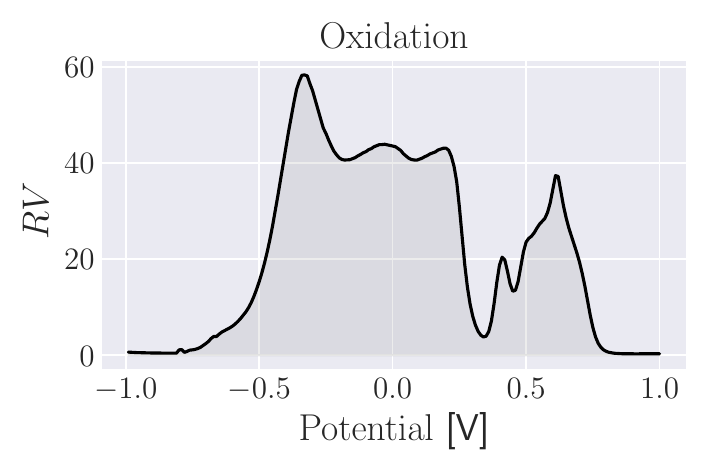}
\end{subfigure}
\begin{subfigure}{0.23\textwidth}
    \centering
    \includegraphics[width=\linewidth]{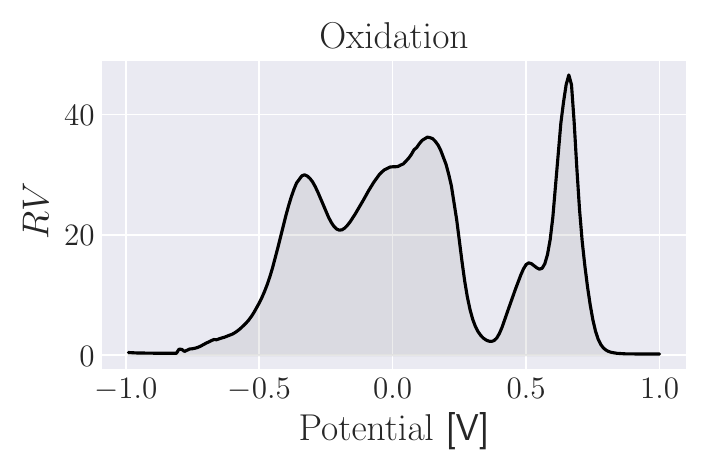}
\end{subfigure}
\begin{subfigure}{0.23\textwidth}
    \centering
    \includegraphics[width=\linewidth]{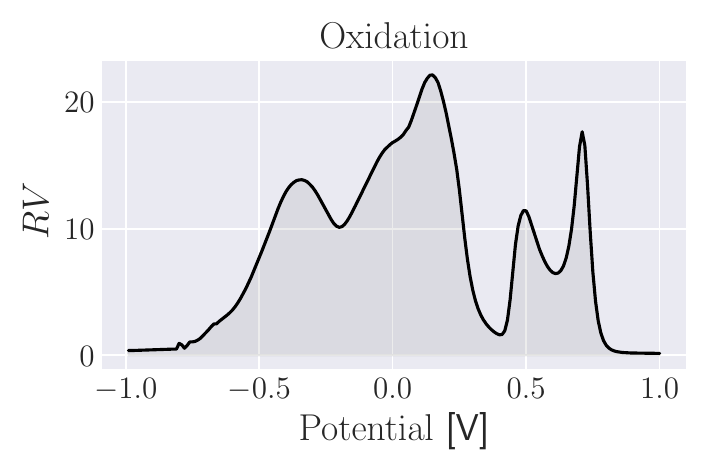}
\end{subfigure}
\begin{subfigure}{0.23\textwidth}
    \centering
    \includegraphics[width=\linewidth]{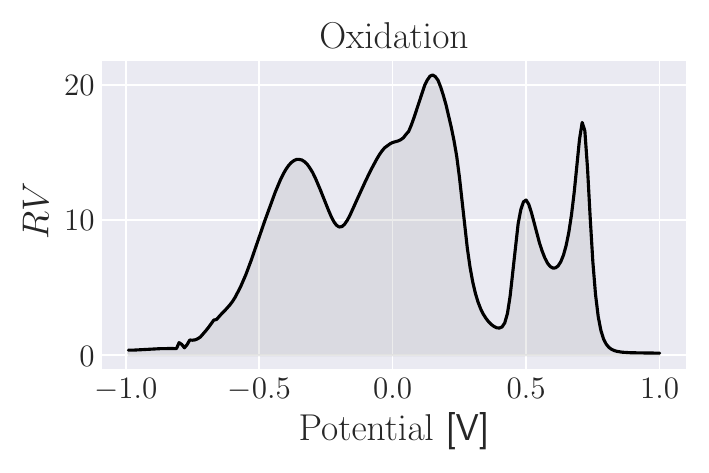}
\end{subfigure}
\\

% ---------------------------------------------------------------------
% OXIDATION: composite score and selected intervals
% ---------------------------------------------------------------------
\begin{subfigure}{0.23\textwidth}
    \centering
    \includegraphics[width=\linewidth]{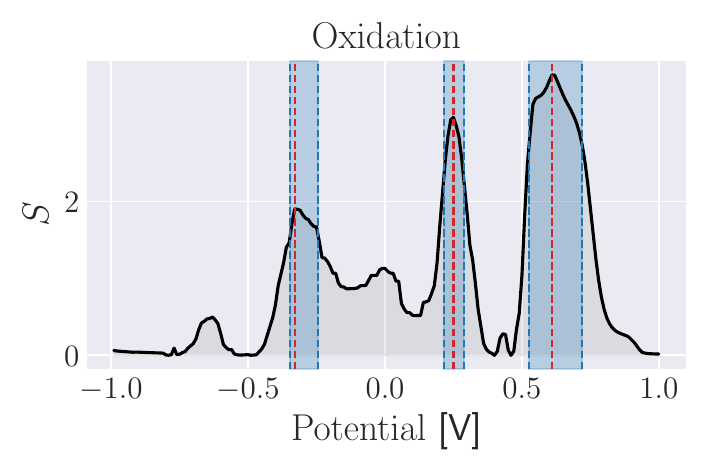}
\end{subfigure}
\begin{subfigure}{0.23\textwidth}
    \centering
    \includegraphics[width=\linewidth]{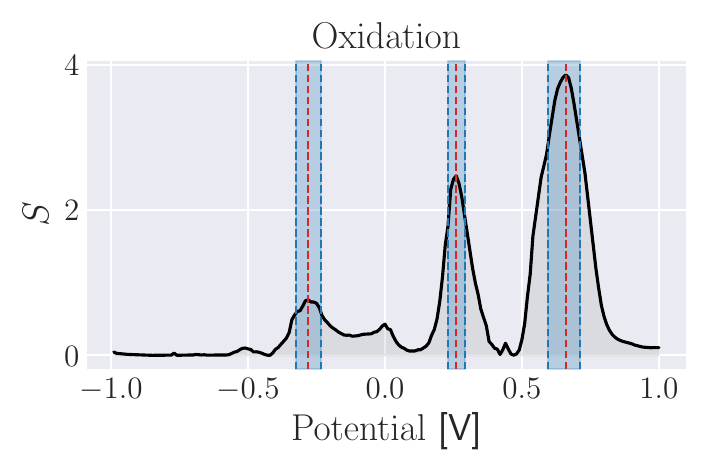}
\end{subfigure}
\begin{subfigure}{0.23\textwidth}
    \centering
    \includegraphics[width=\linewidth]{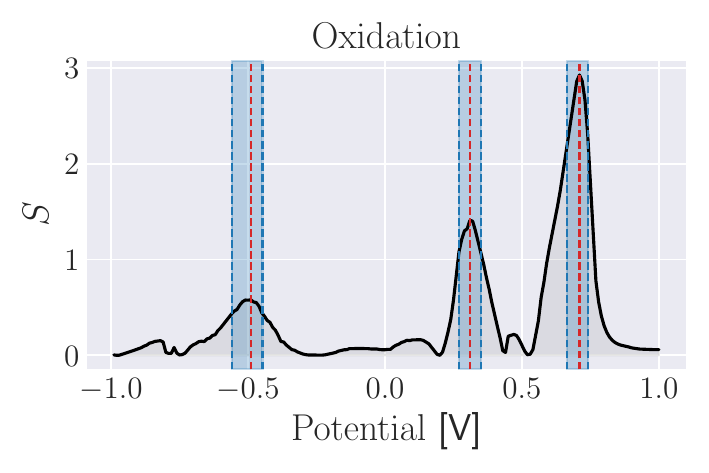}
\end{subfigure}
\begin{subfigure}{0.23\textwidth}
    \centering
    \includegraphics[width=\linewidth]{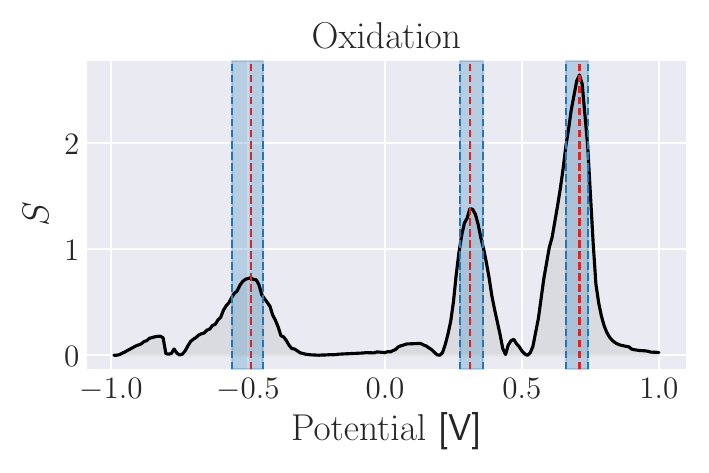}
\end{subfigure}
\\

% ---------------------------------------------------------------------
% REDUCTION: squared Spearman correlation
% ---------------------------------------------------------------------
\begin{subfigure}{0.23\textwidth}
    \centering
    \includegraphics[width=\linewidth]{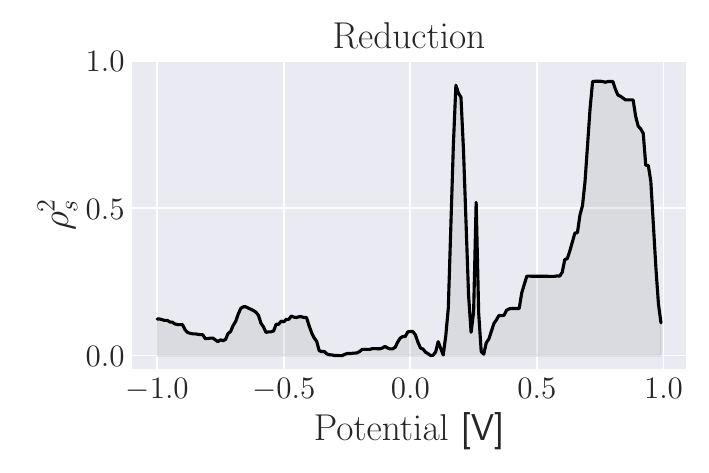}
\end{subfigure}
\begin{subfigure}{0.23\textwidth}
    \centering
    \includegraphics[width=\linewidth]{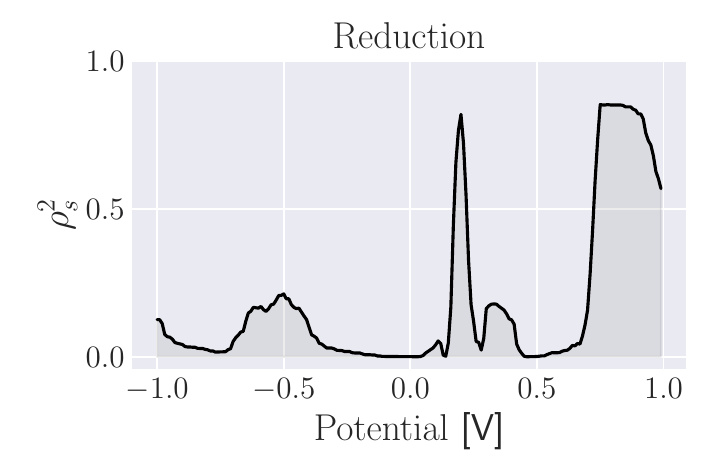}
\end{subfigure}
\begin{subfigure}{0.23\textwidth}
    \centering
    \includegraphics[width=\linewidth]{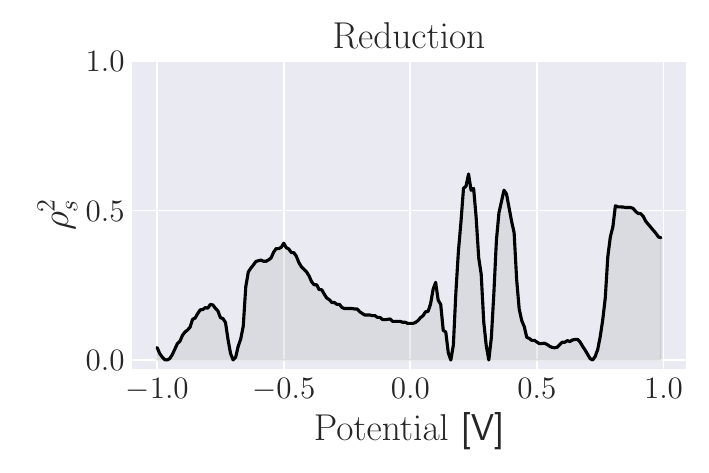}
\end{subfigure}
\begin{subfigure}{0.23\textwidth}
    \centering
    \includegraphics[width=\linewidth]{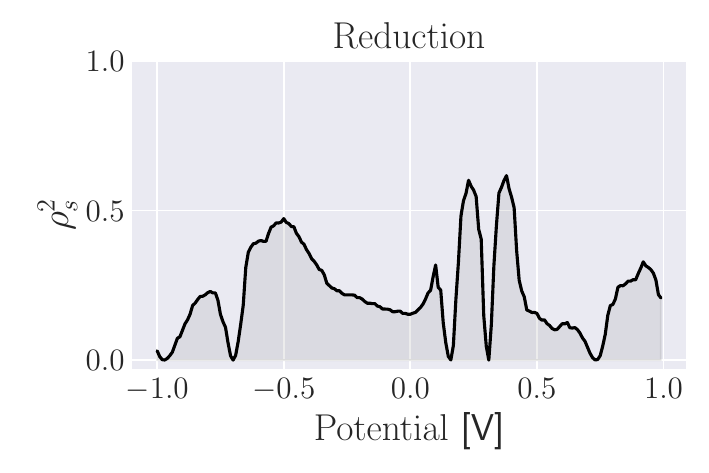}
\end{subfigure}
\\

% ---------------------------------------------------------------------
% REDUCTION: measurement-quality metric
% ---------------------------------------------------------------------
\begin{subfigure}{0.23\textwidth}
    \centering
    \includegraphics[width=\linewidth]{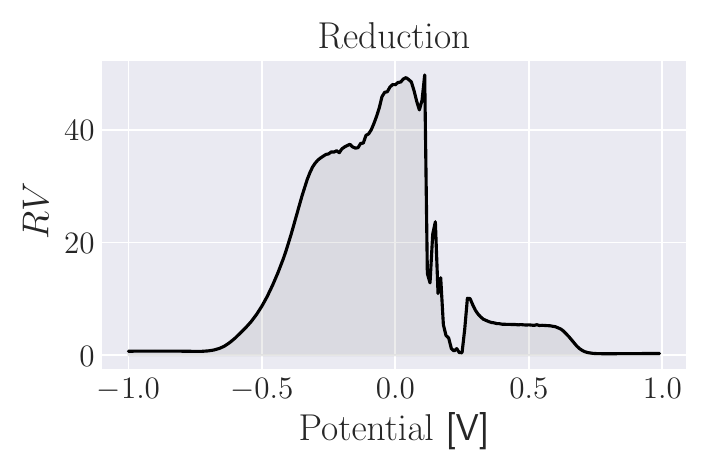}
\end{subfigure}
\begin{subfigure}{0.23\textwidth}
    \centering
    \includegraphics[width=\linewidth]{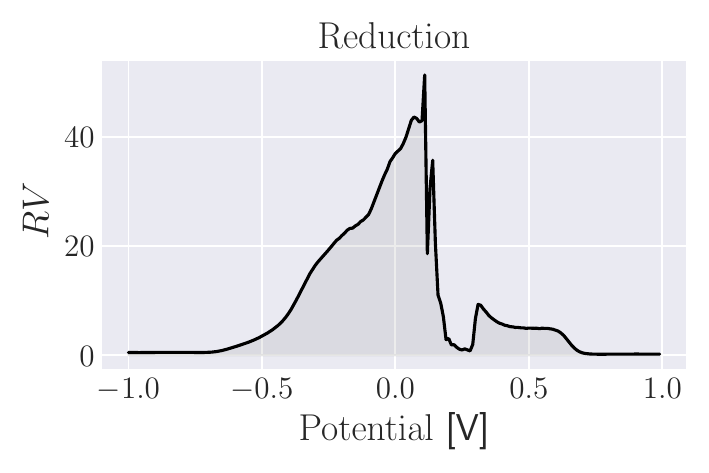}
\end{subfigure}
\begin{subfigure}{0.23\textwidth}
    \centering
    \includegraphics[width=\linewidth]{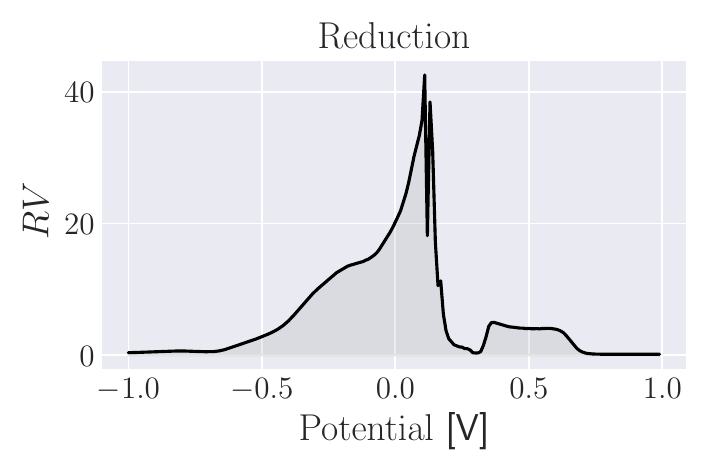}
\end{subfigure}
\begin{subfigure}{0.23\textwidth}
    \centering
    \includegraphics[width=\linewidth]{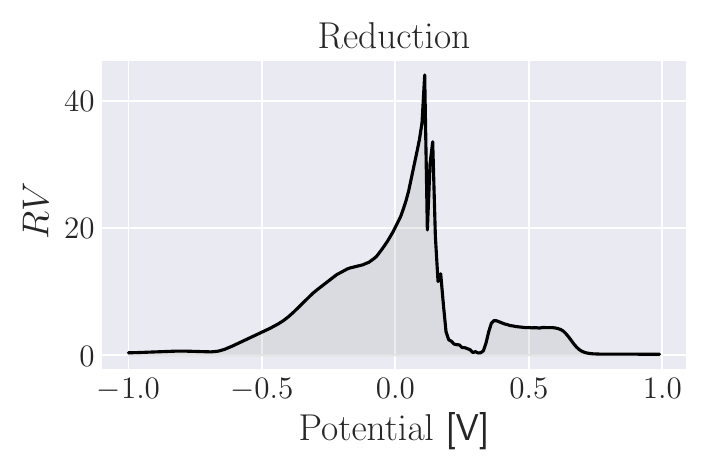}
\end{subfigure}
\\

% ---------------------------------------------------------------------
% REDUCTION: composite score and selected intervals
% ---------------------------------------------------------------------
\begin{subfigure}{0.23\textwidth}
    \centering
    \includegraphics[width=\linewidth]{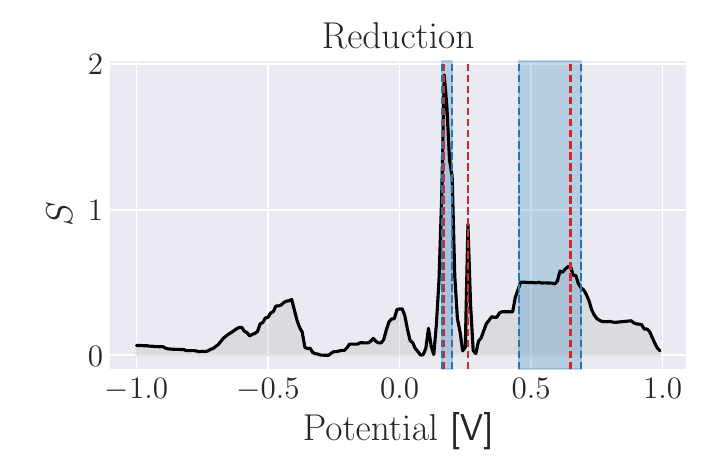}
\end{subfigure}
\begin{subfigure}{0.23\textwidth}
    \centering
    \includegraphics[width=\linewidth]{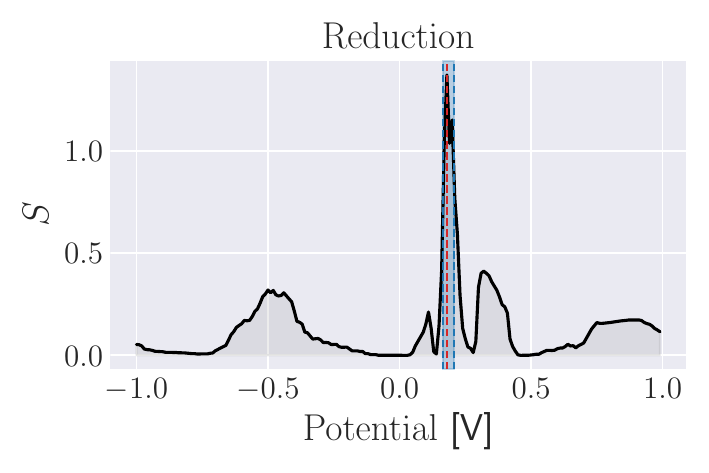}
\end{subfigure}
\begin{subfigure}{0.23\textwidth}
    \centering
    \includegraphics[width=\linewidth]{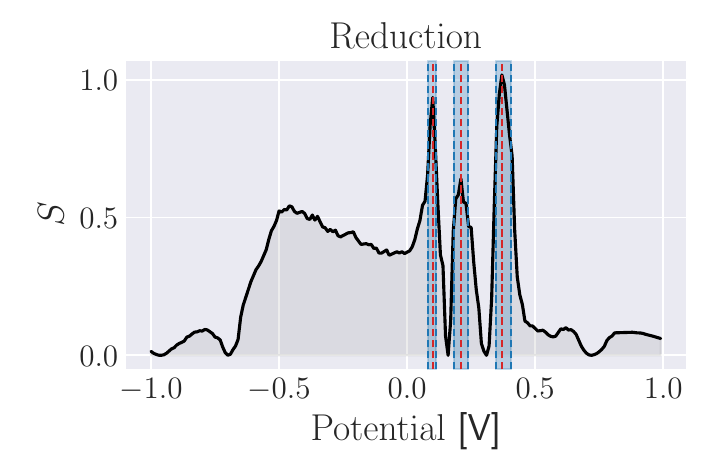}
\end{subfigure}
\begin{subfigure}{0.23\textwidth}
    \centering
    \includegraphics[width=\linewidth]{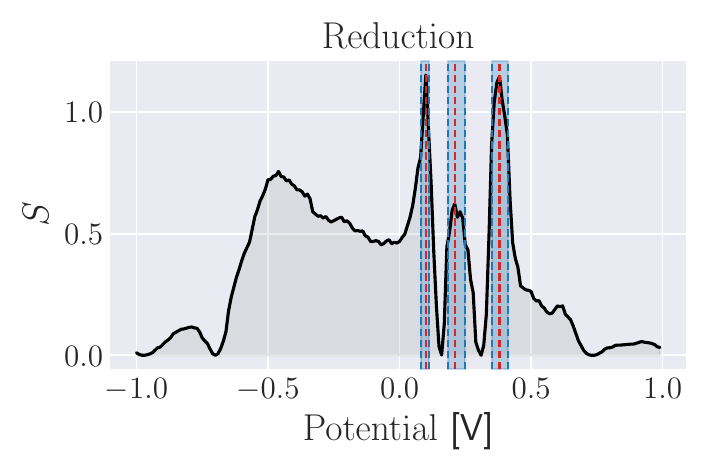}
\end{subfigure}

\caption{Automated candidate feature-region generation for progressively broader nominal glucose concentration ranges: A) 0--75~g/L; B) 0--150~g/L; C) 0--250~g/L; and D) 0--350~g/L. For each concentration range, the upper three rows show the oxidation-channel point-wise metrics: squared Spearman's rank correlation coefficient, measurement-quality metric, and composite information-score fingerprint, respectively. The lower three rows show the corresponding metrics for the reduction channel. All point-wise metrics are calculated using the training set. For all investigated ranges, dominant peaks are detected separately in the oxidation and reduction fingerprints using a shared prominence threshold based on the maximum information score across both branches. Detected peak locations are indicated by red dashed lines. Candidate
intervals are generated from the corresponding peak widths, and the
retained intervals are shown as blue-shaded regions. These retained
intervals define the candidate feature regions subsequently evaluated in the GP train--validation model-selection pipeline.}

\label{fig:scores}
\end{figure*}

For the nominal glucose concentration range of 0--75~g/L, column A of Fig.~\ref{fig:scores} presents the squared Spearman's rank correlation coefficient, the measurement-quality metric, and the resulting composite information fingerprint for the oxidation and reduction channels. The retained candidate intervals are also shown. In total, the information fingerprints yielded five candidate regions: three from the oxidation branch and two from the reduction branch. Evaluating all non-empty combinations of these regions resulted in $2^5-1=31$ candidate GPR models. 

Validation-based ranking selected a combination comprising the oxidation region from 0.53 to 0.72~V and the reduction region from 0.45 to 0.69~V, as summarised in Table~\ref{tab:final_models}. Together, these regions contain 42 of the original 400 measurement channels, representing an 89.5\% reduction in input dimensionality. Interestingly, neither selected region coincides with the dominant current-magnitude peaks visible in the original voltammograms. This demonstrates that the proposed methodology can identify informative regions that are not immediately evident from visual inspection of the raw measurements. Moreover, although the oxidation region had the highest average fingerprint score among the retained candidates, the selected reduction region had the lowest. This result illustrates the nontrivial complementary roles of the two framework stages; that is, the information fingerprint generates data-informed candidate regions, whereas validation-based model selection identifies combinations with joint predictive value.

\begin{table*}[htb!]
\centering
\caption{Summary of the selected feature-region soft sensors and comparison
with the corresponding full-feature benchmarks across the investigated nominal
glucose concentration ranges. Ox: oxidation branch. Red: reduction branch.}
\label{tab:final_models}

\footnotesize
\setlength{\tabcolsep}{4pt}
\renewcommand{\arraystretch}{1.05}

\begin{tabular}{
    @{}
    >{\raggedright\arraybackslash}p{0.07\textwidth}
    >{\raggedright\arraybackslash}p{0.32\textwidth}
    >{\centering\arraybackslash}p{0.08\textwidth}
    >{\centering\arraybackslash}p{0.10\textwidth}
    >{\centering\arraybackslash}p{0.10\textwidth}
    >{\centering\arraybackslash}p{0.13\textwidth}
    @{}
}

\toprule

\multirow[c]{2}{0.07\textwidth}{\centering\textbf{Range}\\\textbf{[g/L]}} &
\multirow[c]{2}{0.32\textwidth}{\centering\textbf{Selected potential regions [V]}} &
\multirow[c]{2}{0.08\textwidth}{\centering\textbf{Channels}\\\textbf{(regions)}} &
\multicolumn{2}{c}{\textbf{Test RMSE [g/L]}} &
\multirow[c]{2}{0.11\textwidth}{\centering\textbf{RMSE reduction [\%]}} \\

\cmidrule(lr){4-5}

&
&
&
\textbf{Selected} &
\textbf{Full-feature} &
\\

\midrule

0--75 &
Ox: [0.53, 0.72]; Red: [0.45, 0.69] &
42 (2) &
1.84 &
15.70 &
88.3 \\[1mm]

0--150 &
Ox: [0.60, 0.71] &
12 (1) &
2.38 &
38.94 &
93.9 \\[1mm]

0--250 &
Ox: [0.66, 0.74]; Red: [0.08, 0.11] &
11 (2) &
4.09 &
75.64 &
94.6 \\[1mm]

0--350 &
Ox: [0.66, 0.74]; Ox: [0.27, 0.36]; Red: [0.35, 0.41] &
22 (3) &
9.98 &
101.66 &
90.2 \\

\bottomrule
\end{tabular}
\end{table*}

\begin{figure*}[htb!]
\centering

% =====================================================================
% Column headings
% =====================================================================
\makebox[\textwidth][c]{%
\makebox[0.23\textwidth][c]{\textbf{A) 0--75 g/L}}%
\makebox[0.23\textwidth][c]{\textbf{B) 0--150 g/L}}%
\makebox[0.23\textwidth][c]{\textbf{C) 0--250 g/L}}%
\makebox[0.23\textwidth][c]{\textbf{D) 0--350 g/L}}%
}

\vspace{1mm}
\makebox[\textwidth][c]{%
\rule{0.29\textwidth}{0.4pt}\hspace{6pt}%
\textbf{\small{\textit{Selected feature-region GPR}}}%
\hspace{6pt}\rule{0.29\textwidth}{0.4pt}}
% \vspace{1mm}

% =====================================================================
% Candidate selected model: TRAINING
% =====================================================================
\begin{subfigure}{0.23\textwidth}
    \centering
    \includegraphics[width=\linewidth]
    {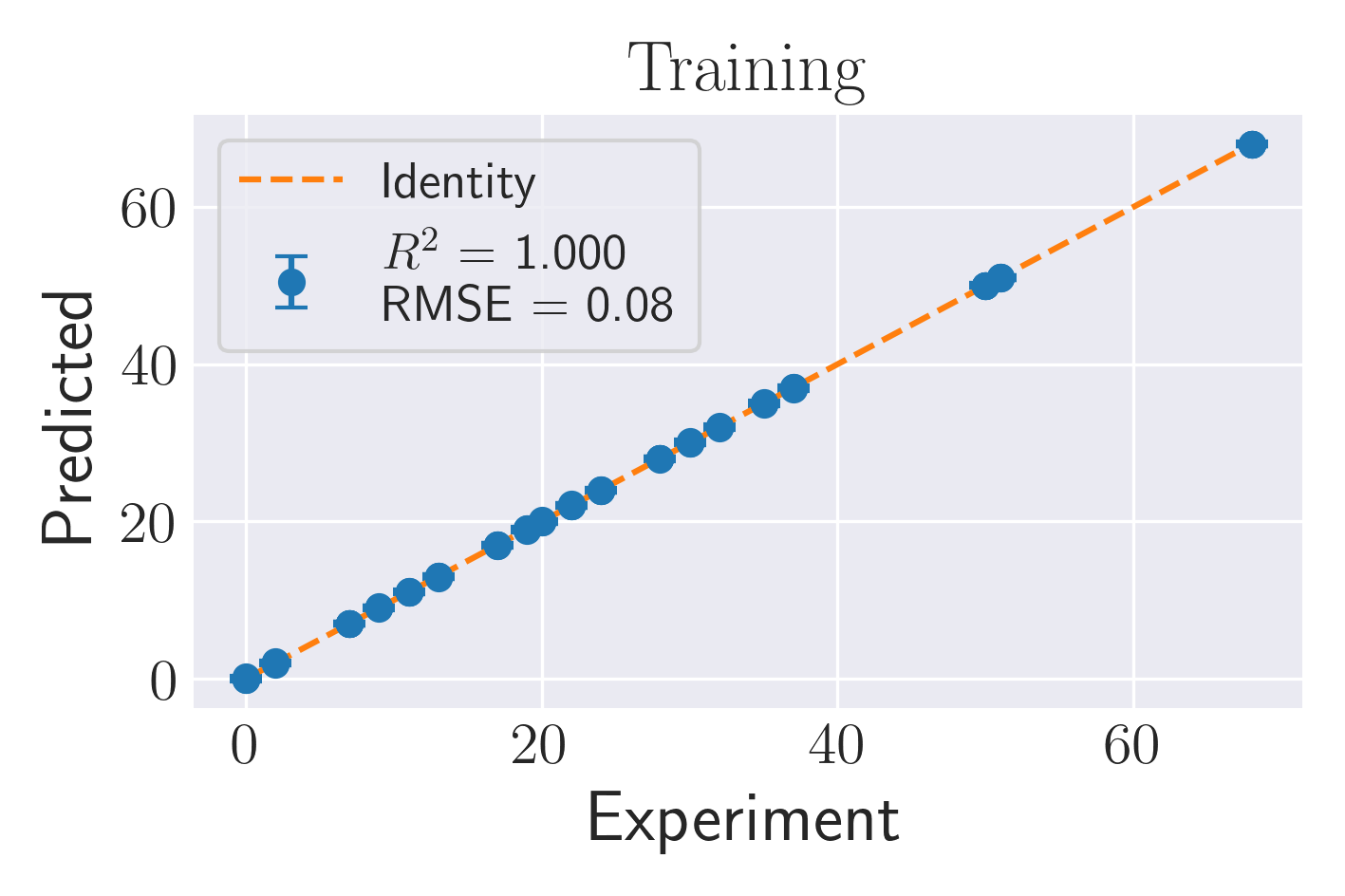}
\end{subfigure}%
\begin{subfigure}{0.23\textwidth}
    \centering
    \includegraphics[width=\linewidth]
    {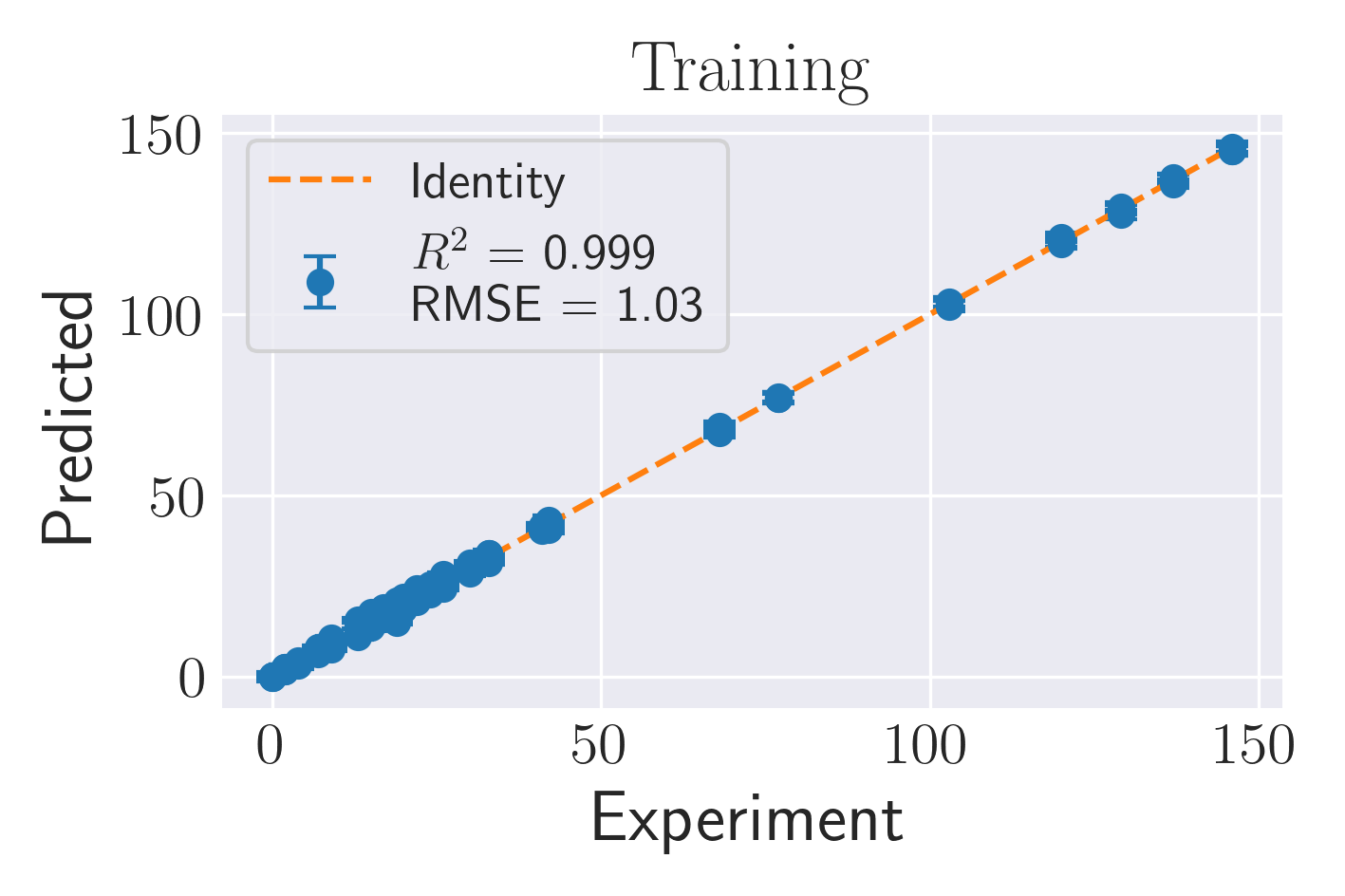}
\end{subfigure}%
\begin{subfigure}{0.23\textwidth}
    \centering
    \includegraphics[width=\linewidth]
    {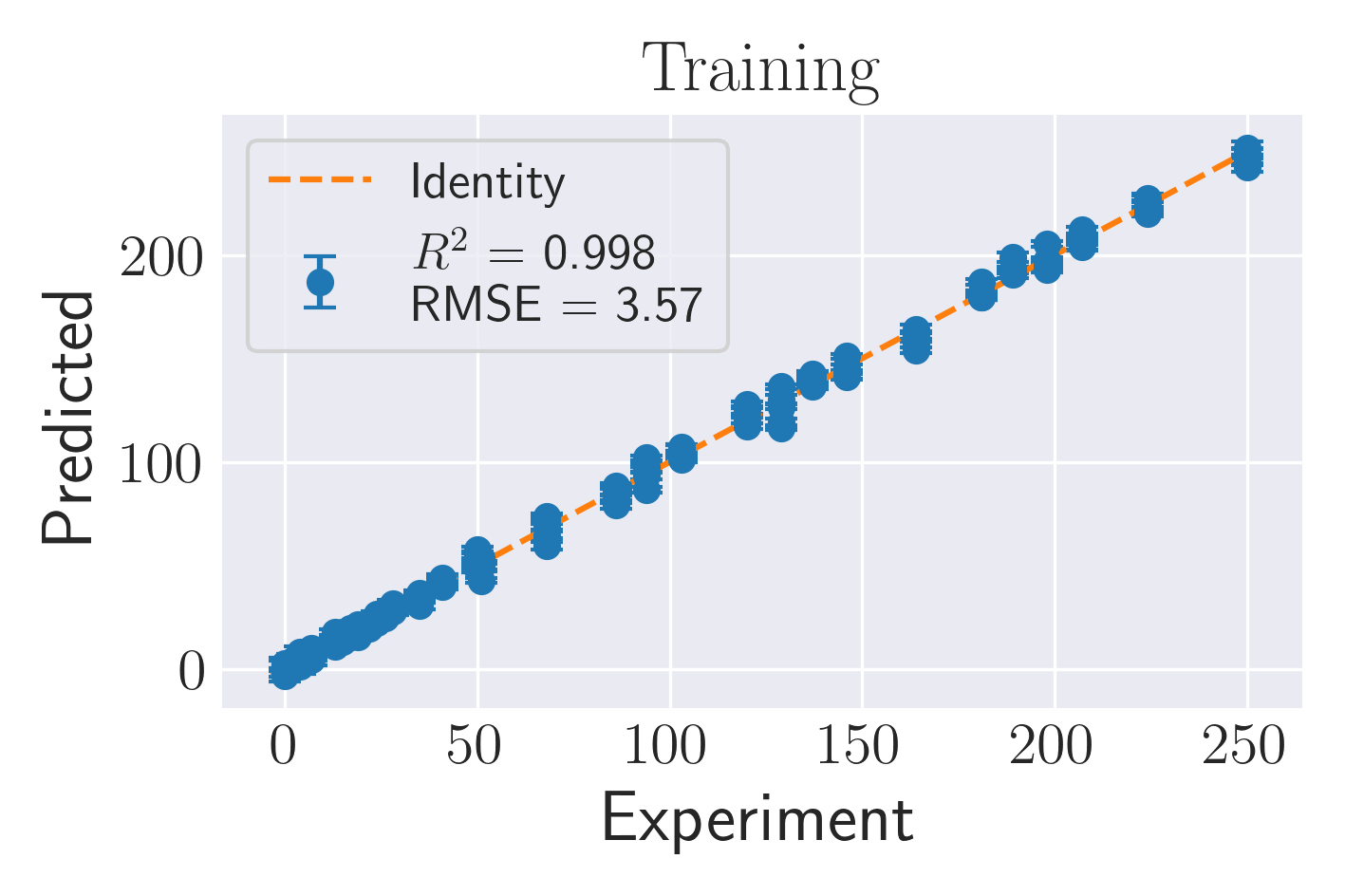}
\end{subfigure}%
\begin{subfigure}{0.23\textwidth}
    \centering
    \includegraphics[width=\linewidth]
    {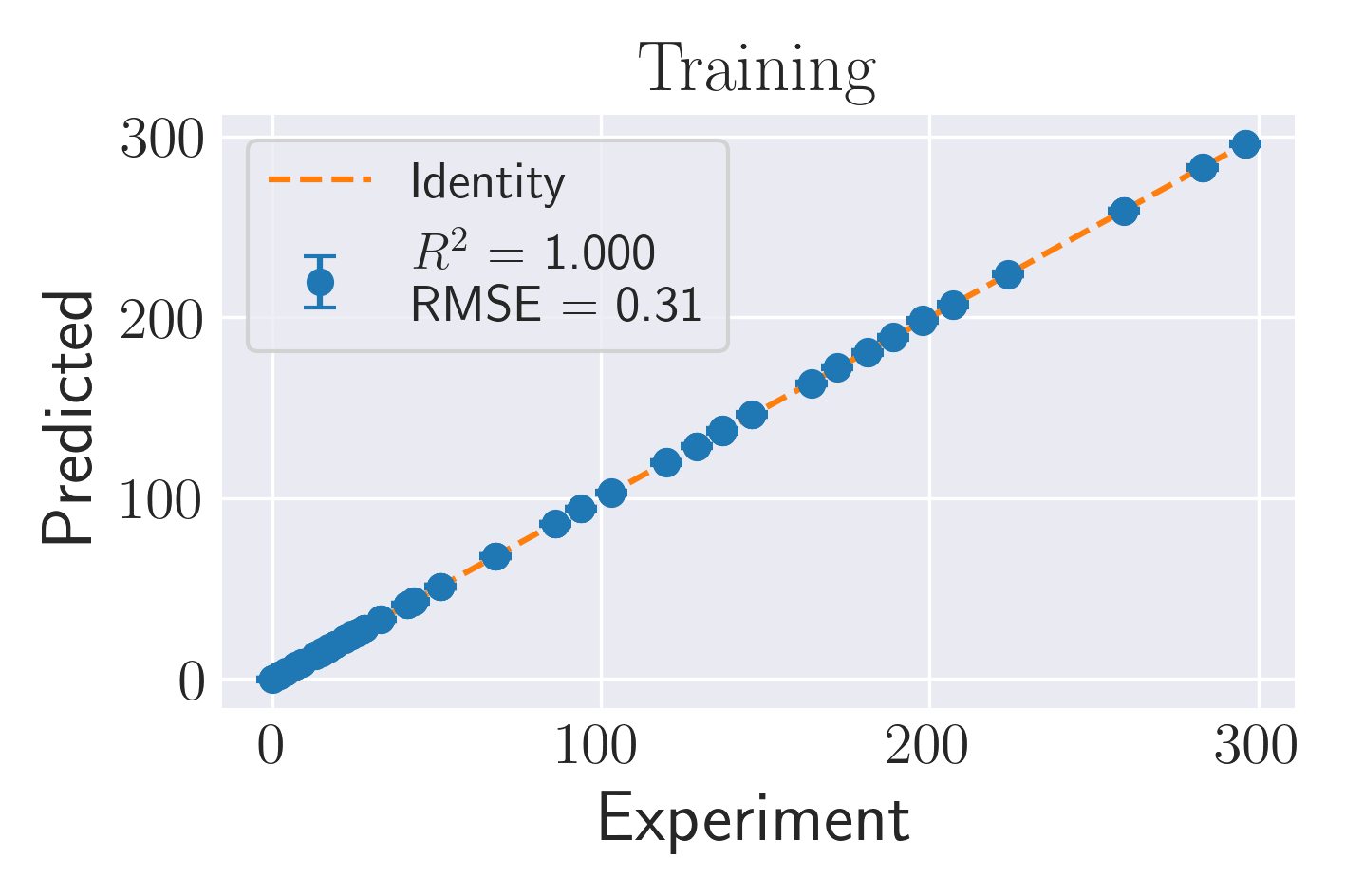}
\end{subfigure}

% =====================================================================
% Candidate selected model: VALIDATION
% =====================================================================
\begin{subfigure}{0.23\textwidth}
    \centering
    \includegraphics[width=\linewidth]
    {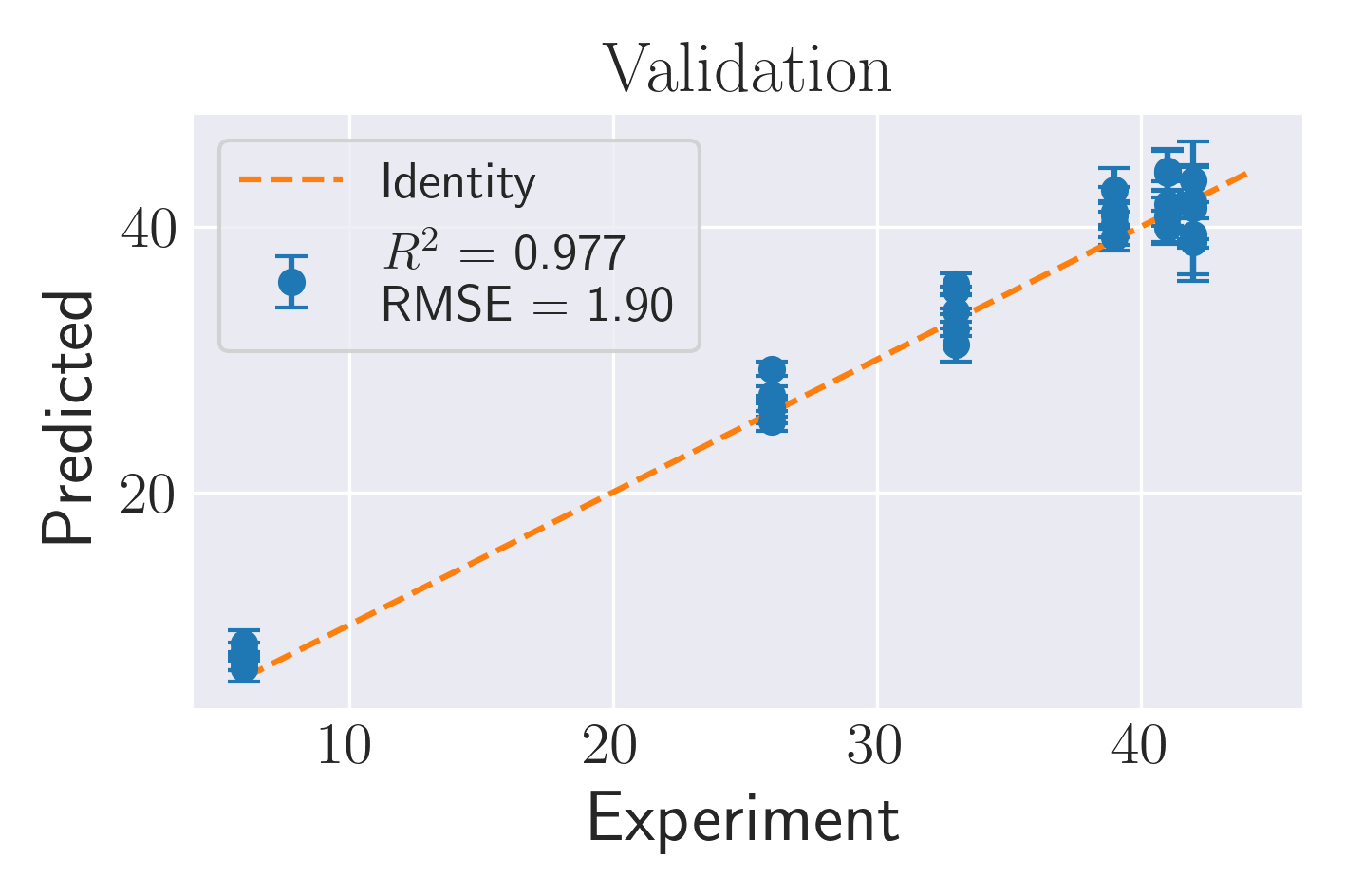}
\end{subfigure}%
\begin{subfigure}{0.23\textwidth}
    \centering
    \includegraphics[width=\linewidth]
    {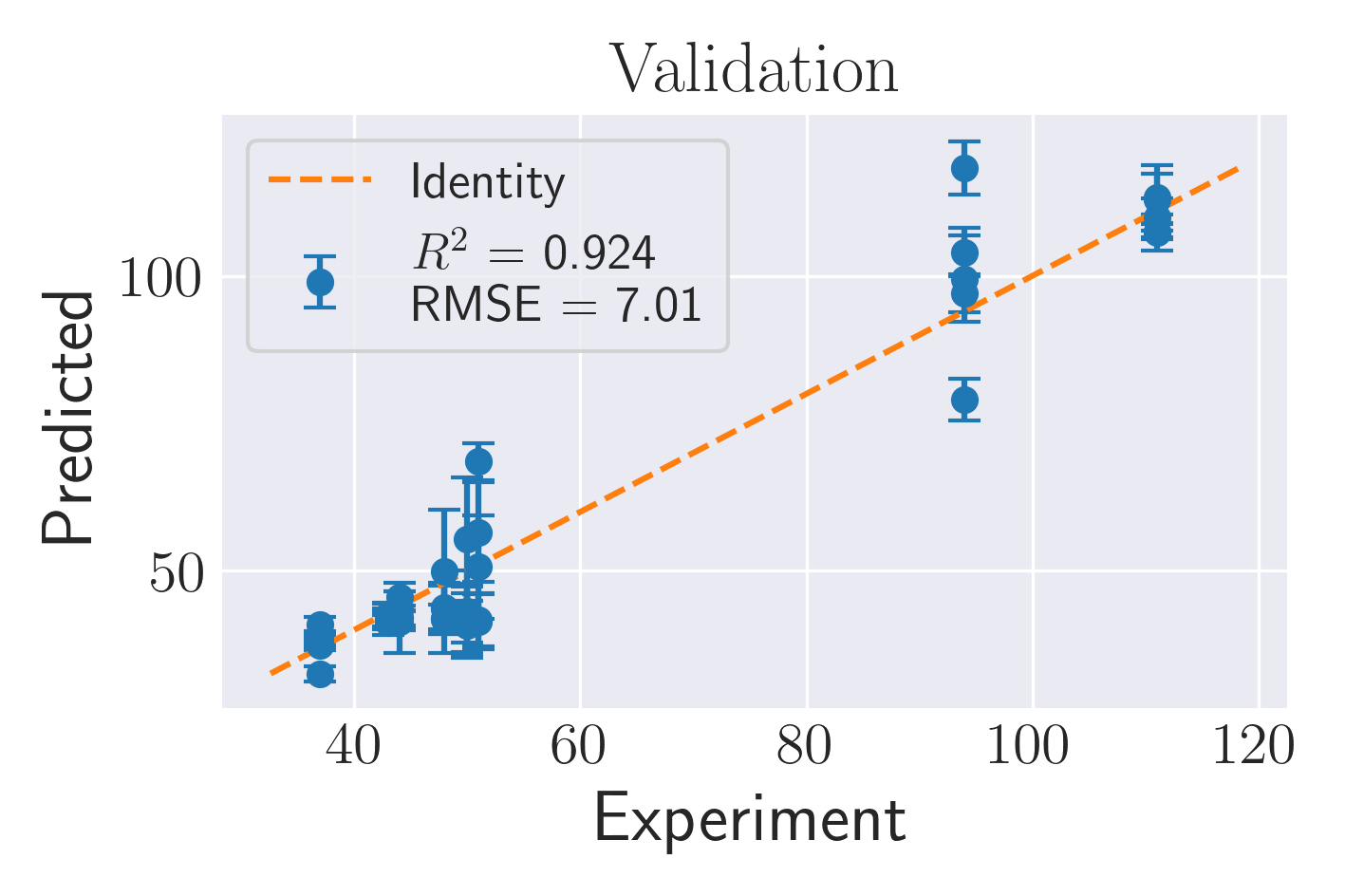}
\end{subfigure}%
\begin{subfigure}{0.23\textwidth}
    \centering
    \includegraphics[width=\linewidth]
    {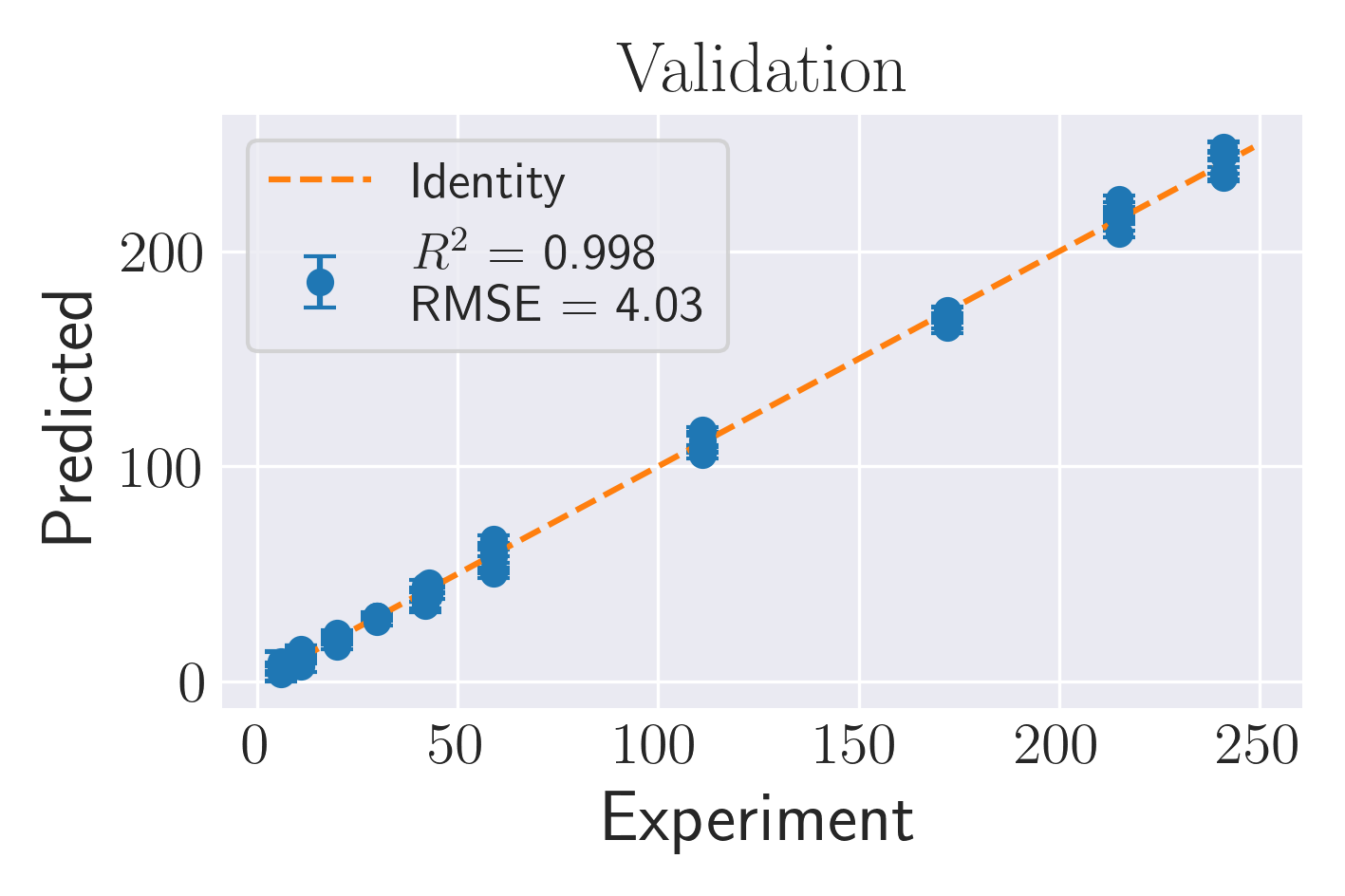}
\end{subfigure}%
\begin{subfigure}{0.23\textwidth}
    \centering
    \includegraphics[width=\linewidth]
    {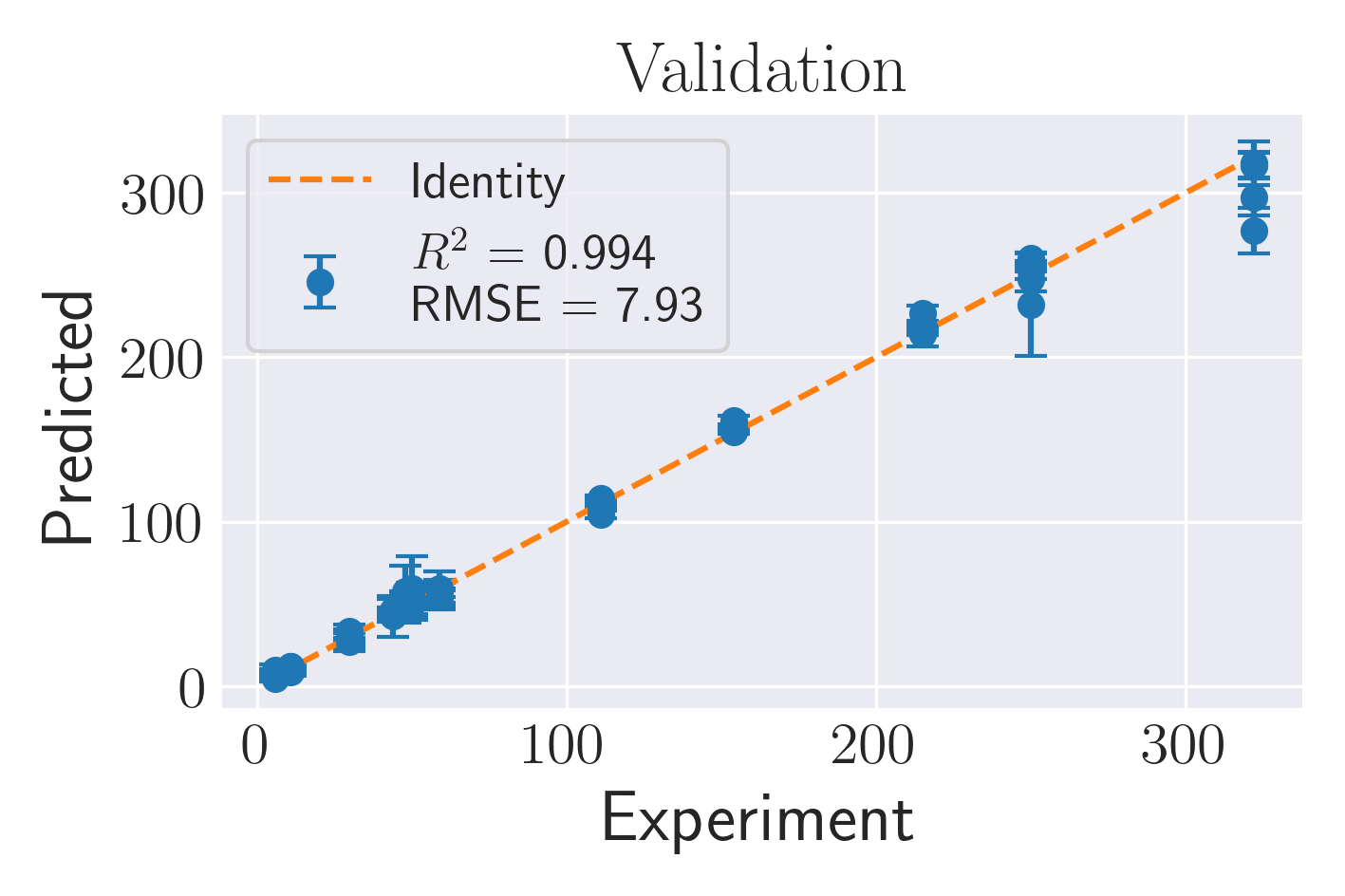}
\end{subfigure}

% =====================================================================
% FINAL REDUCED MODEL: TRAINING + VALIDATION
% =====================================================================
\begin{subfigure}{0.23\textwidth}
    \centering
    \includegraphics[width=\linewidth]
    {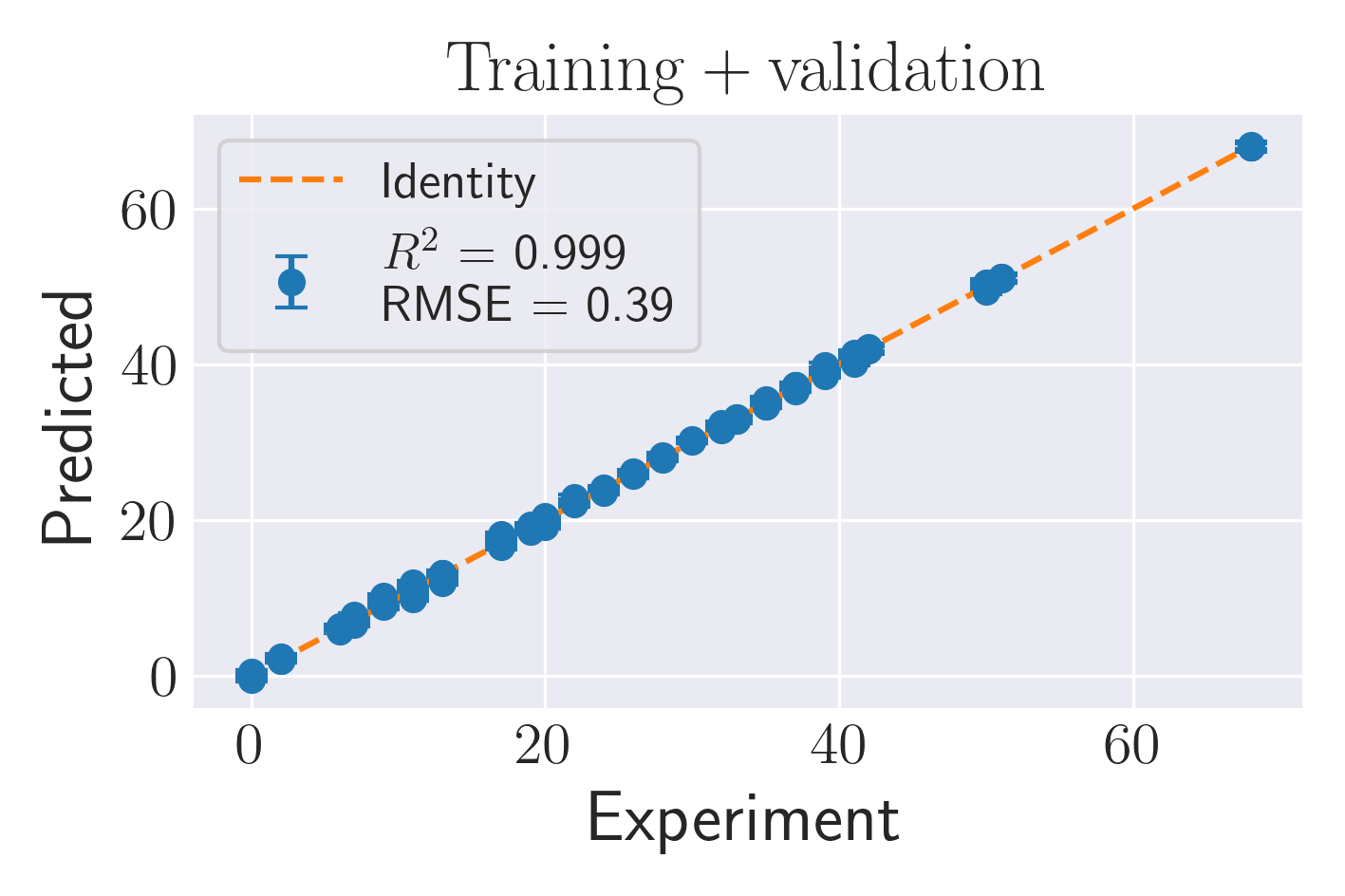}
\end{subfigure}%
\begin{subfigure}{0.23\textwidth}
    \centering
    \includegraphics[width=\linewidth]
    {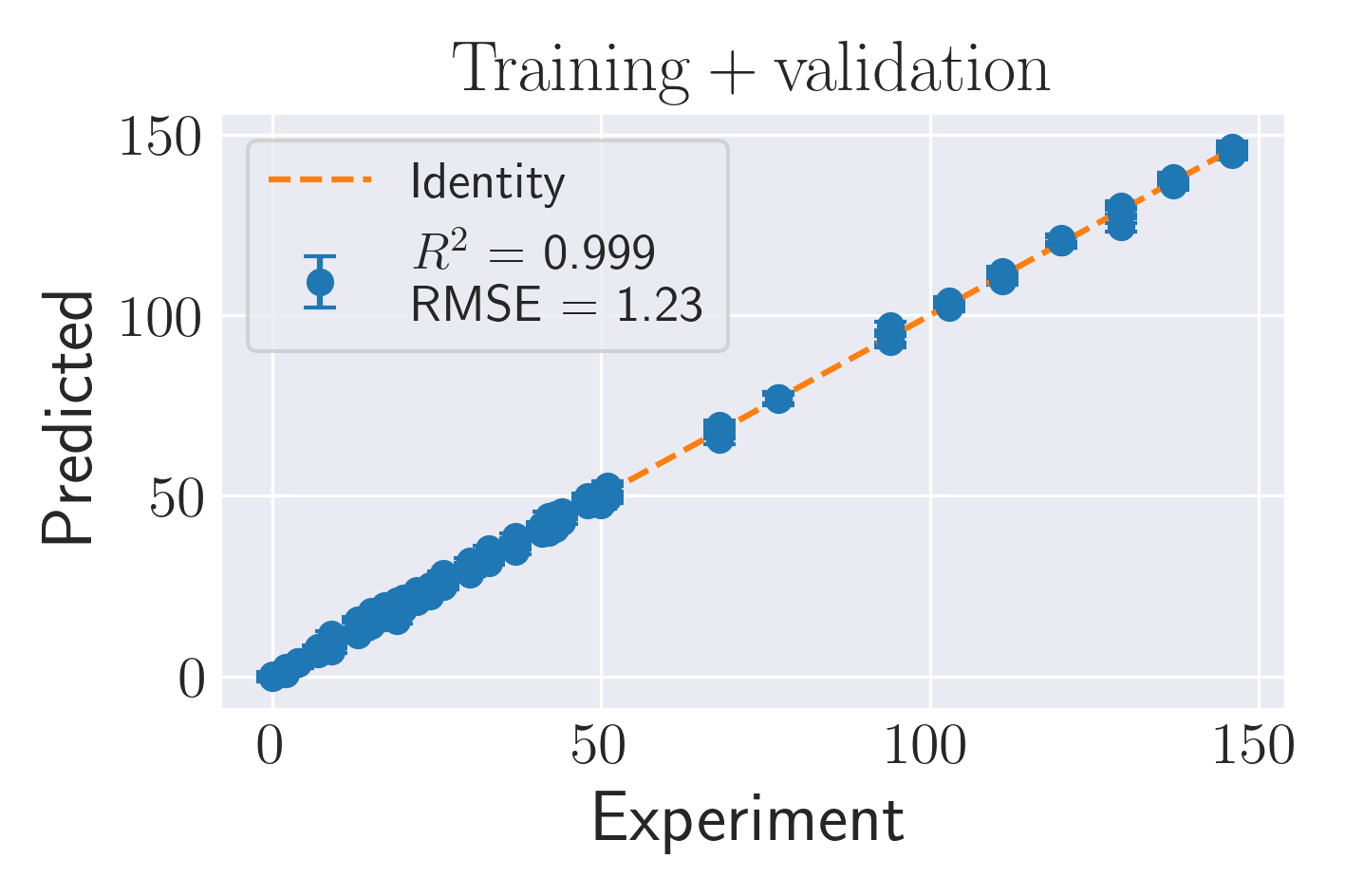}
\end{subfigure}%
\begin{subfigure}{0.23\textwidth}
    \centering
    \includegraphics[width=\linewidth]
    {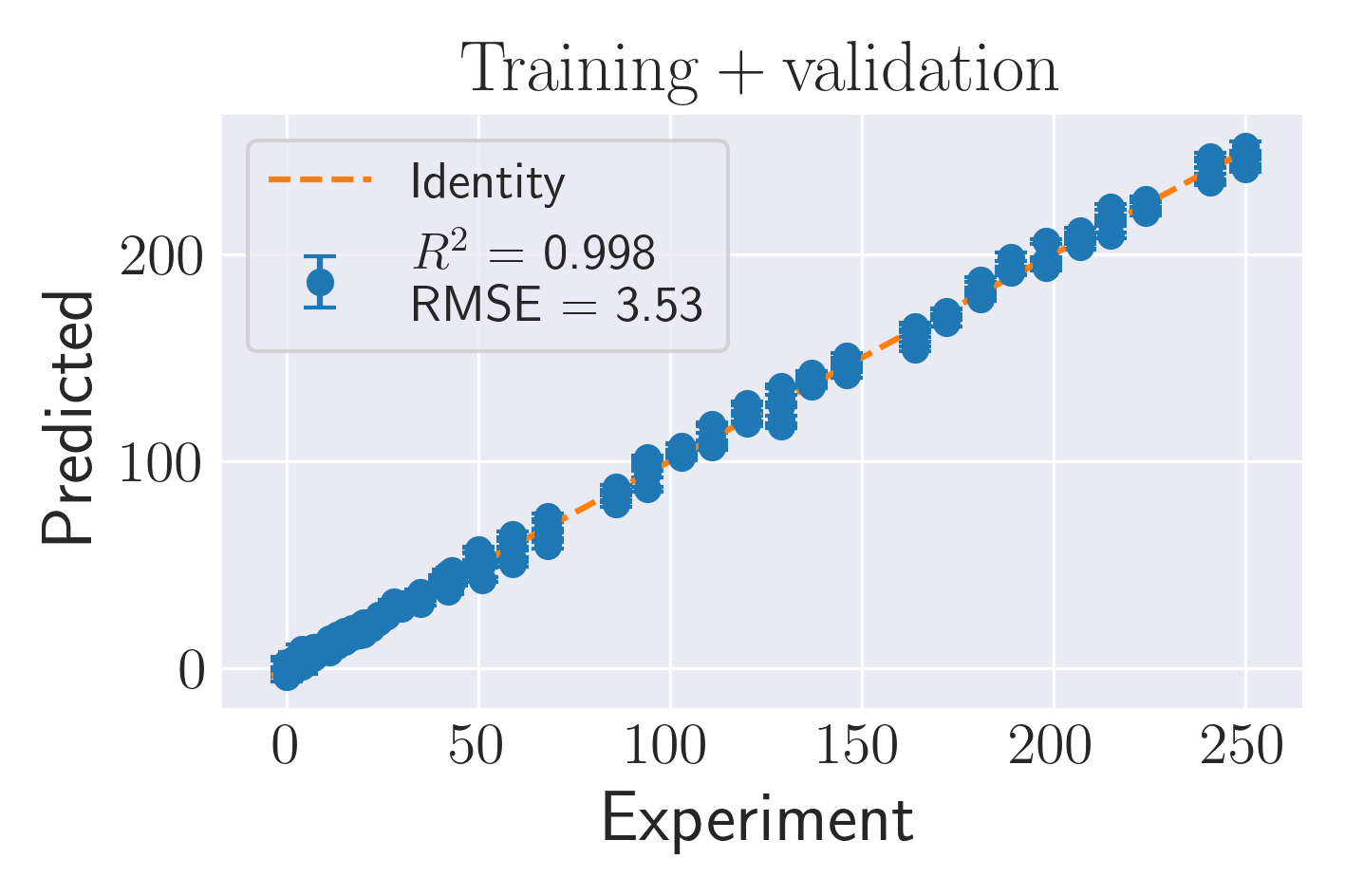}
\end{subfigure}%
\begin{subfigure}{0.23\textwidth}
    \centering
    \includegraphics[width=\linewidth]
    {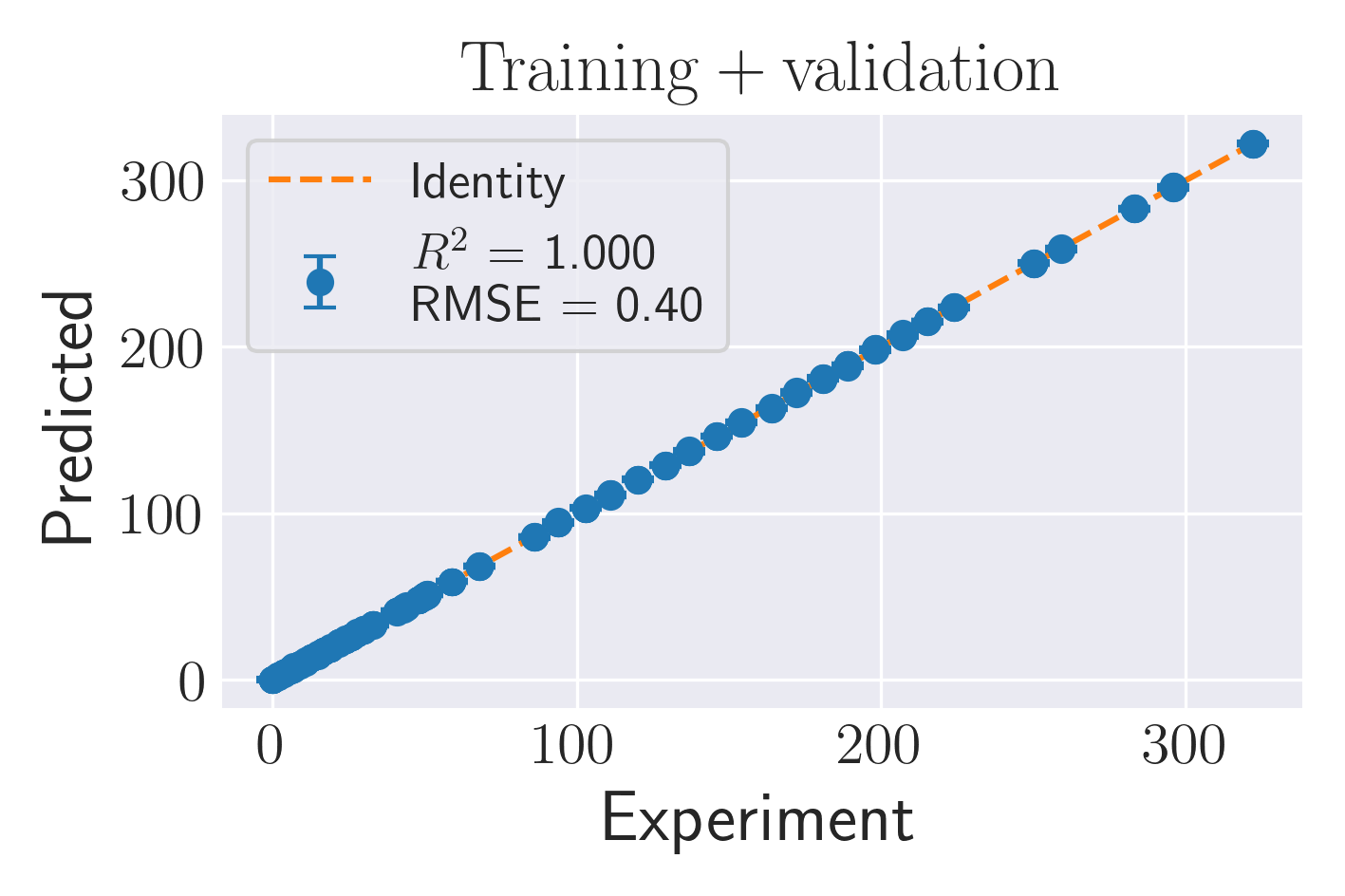}
\end{subfigure}

% =====================================================================
% FINAL REDUCED MODEL: TEST
% =====================================================================
\begin{subfigure}{0.23\textwidth}
    \centering
    \includegraphics[width=\linewidth]
    {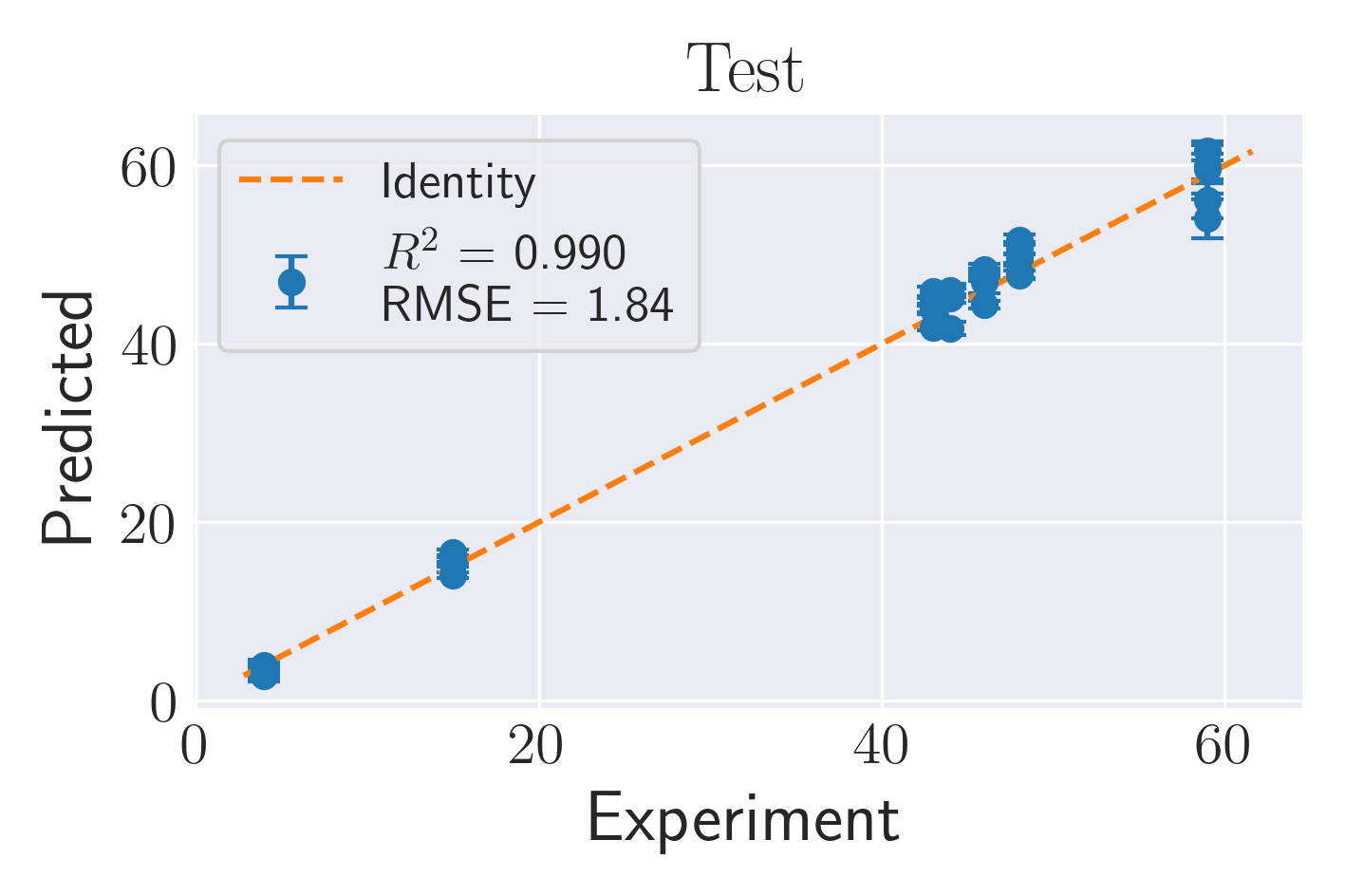}
\end{subfigure}%
\begin{subfigure}{0.23\textwidth}
    \centering
    \includegraphics[width=\linewidth]
    {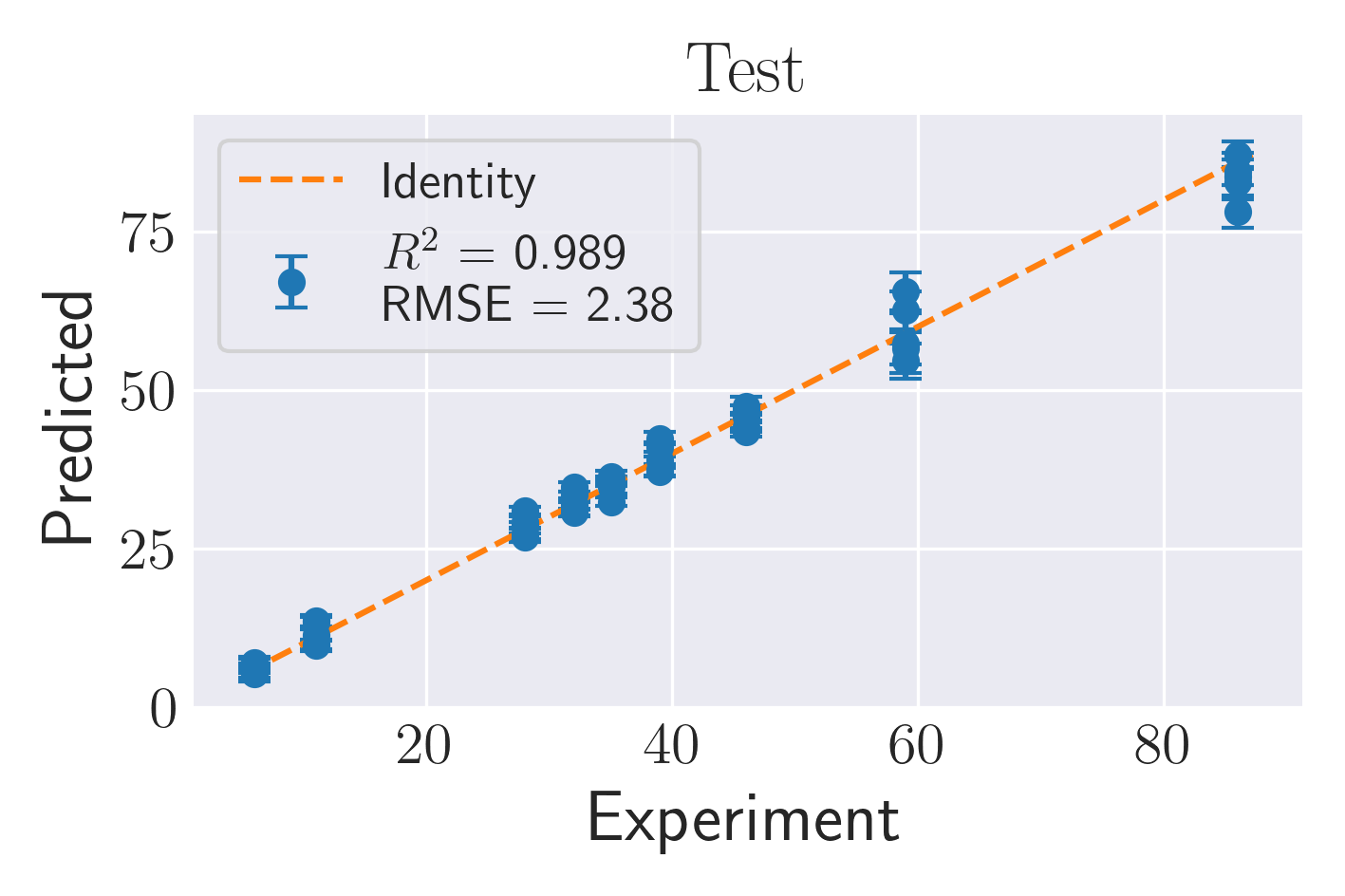}
\end{subfigure}%
\begin{subfigure}{0.23\textwidth}
    \centering
    \includegraphics[width=\linewidth]
    {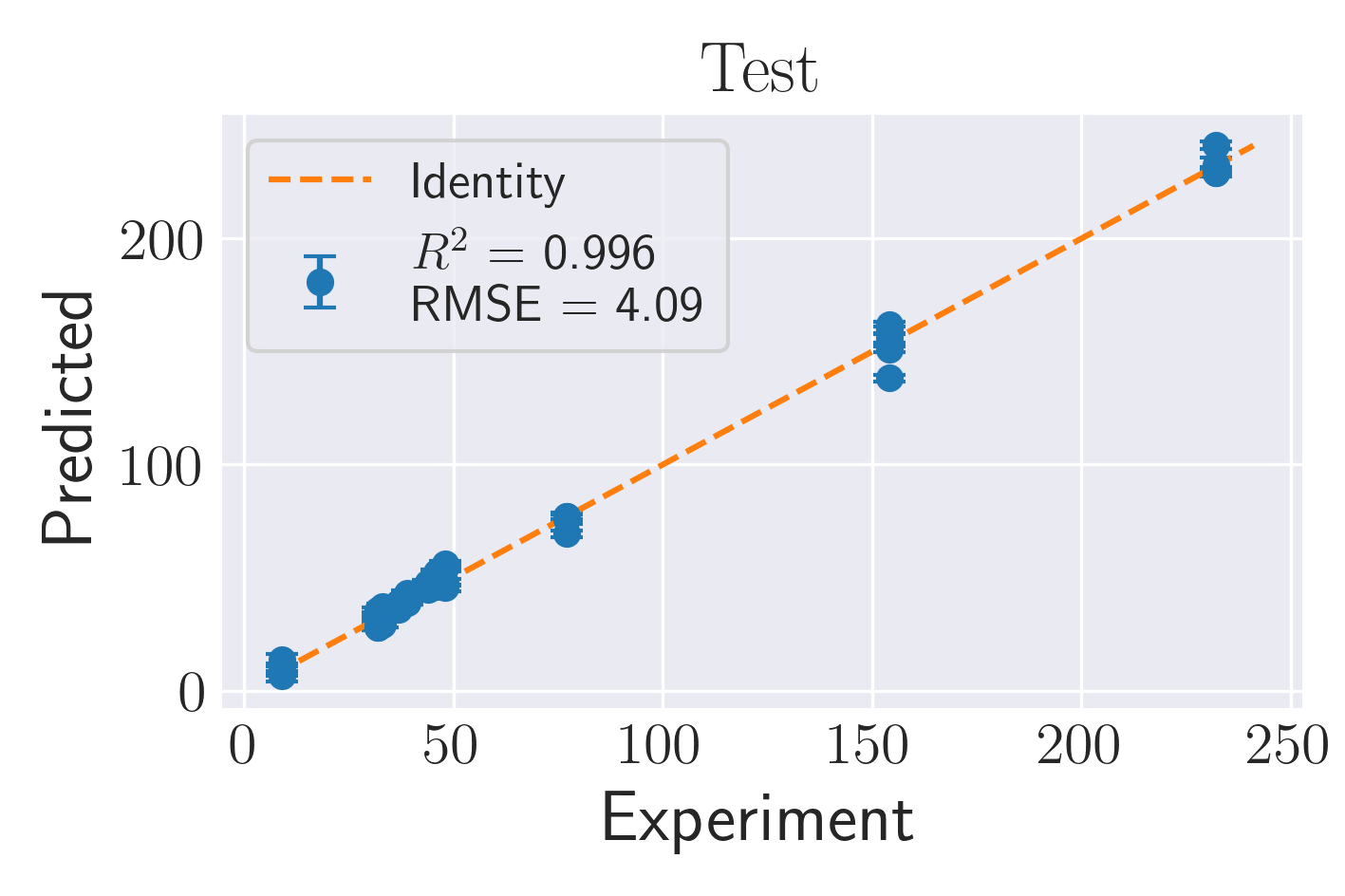}
\end{subfigure}%
\begin{subfigure}{0.23\textwidth}
    \centering
    \includegraphics[width=\linewidth]
    {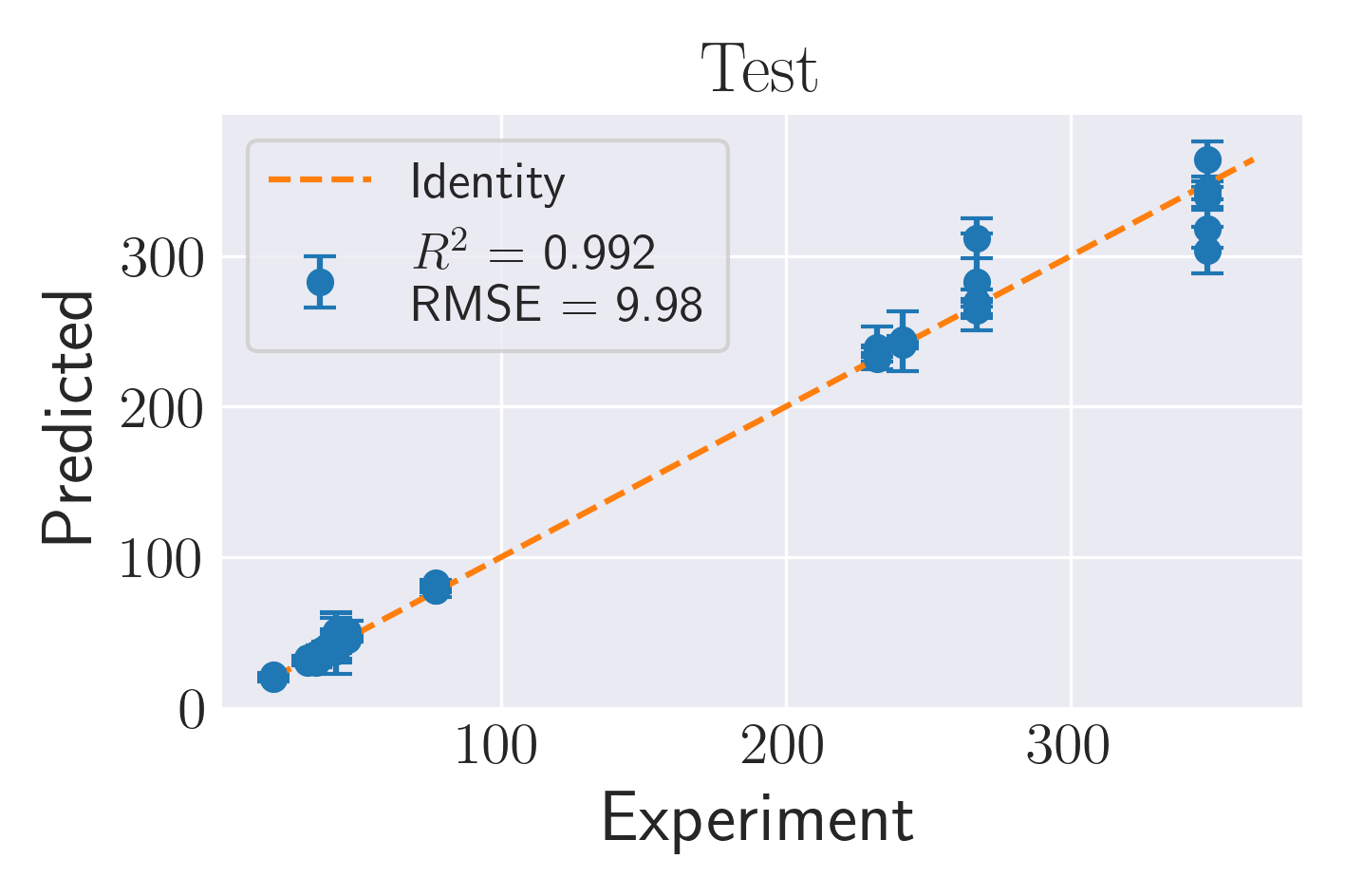}
\end{subfigure}

% \vspace{-1mm}
\makebox[\textwidth][c]{%
\rule{0.31\textwidth}{0.4pt}\hspace{6pt}%
\textbf{\small{\textit{Full-feature GPR benchmark}}}%
\hspace{6pt}\rule{0.31\textwidth}{0.4pt}}
\vspace{1mm}
% =====================================================================
% FULL-FEATURE MODEL: TRAINING + VALIDATION
% =====================================================================
\begin{subfigure}{0.23\textwidth}
    \centering
    \includegraphics[width=\linewidth]
    {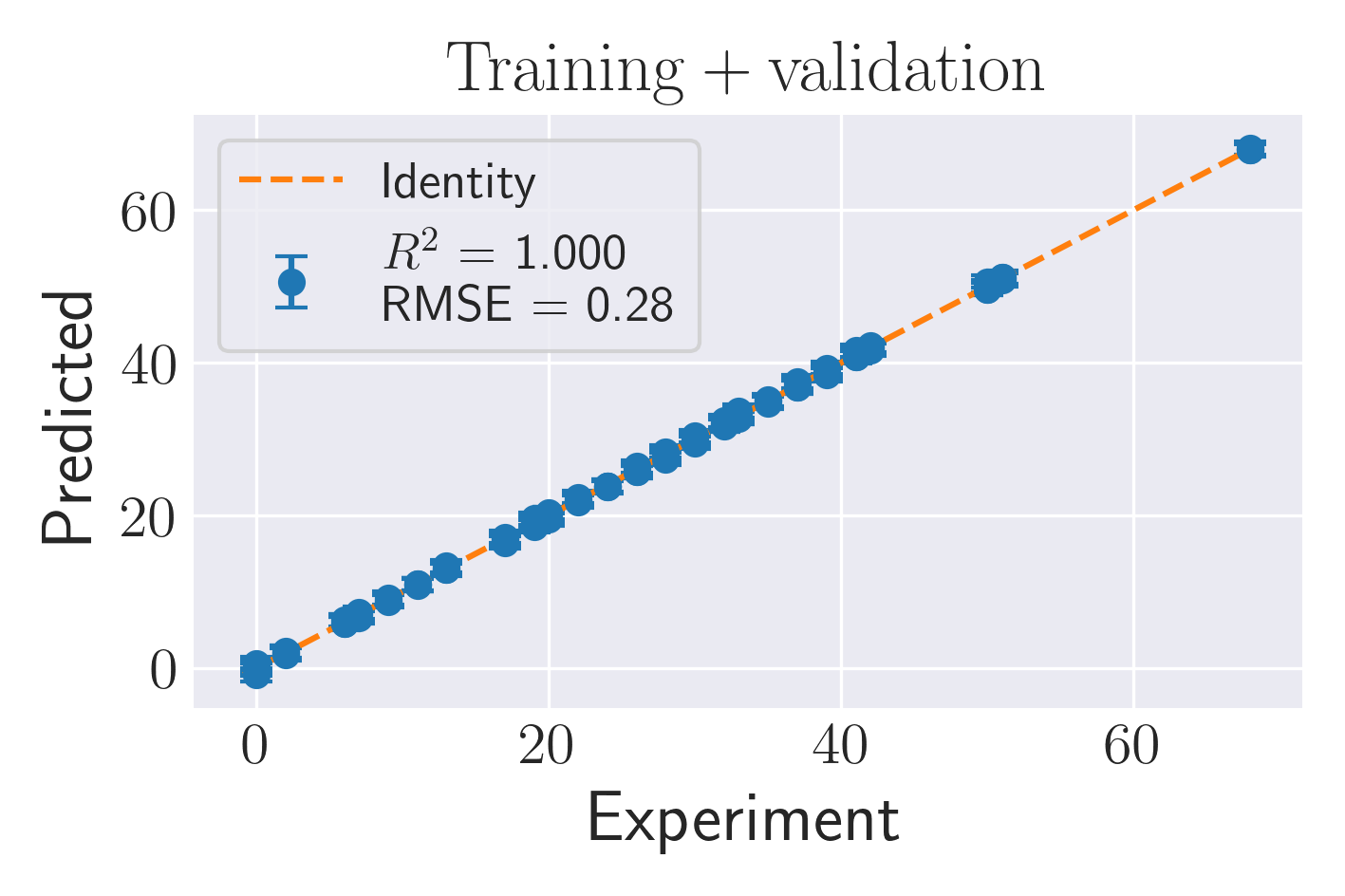}
\end{subfigure}%
\begin{subfigure}{0.23\textwidth}
    \centering
    \includegraphics[width=\linewidth]
    {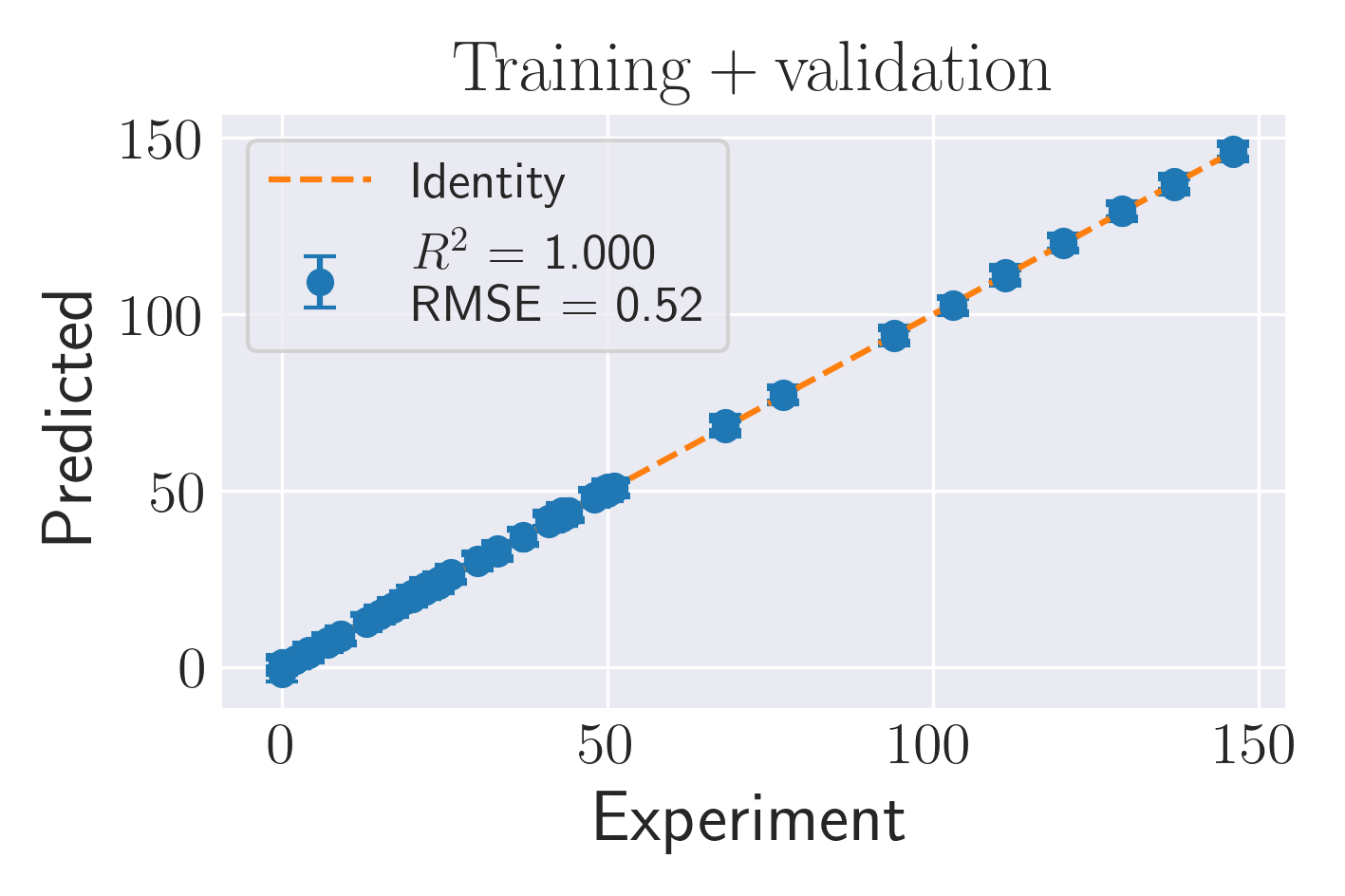}
\end{subfigure}%
\begin{subfigure}{0.23\textwidth}
    \centering
    \includegraphics[width=\linewidth]
    {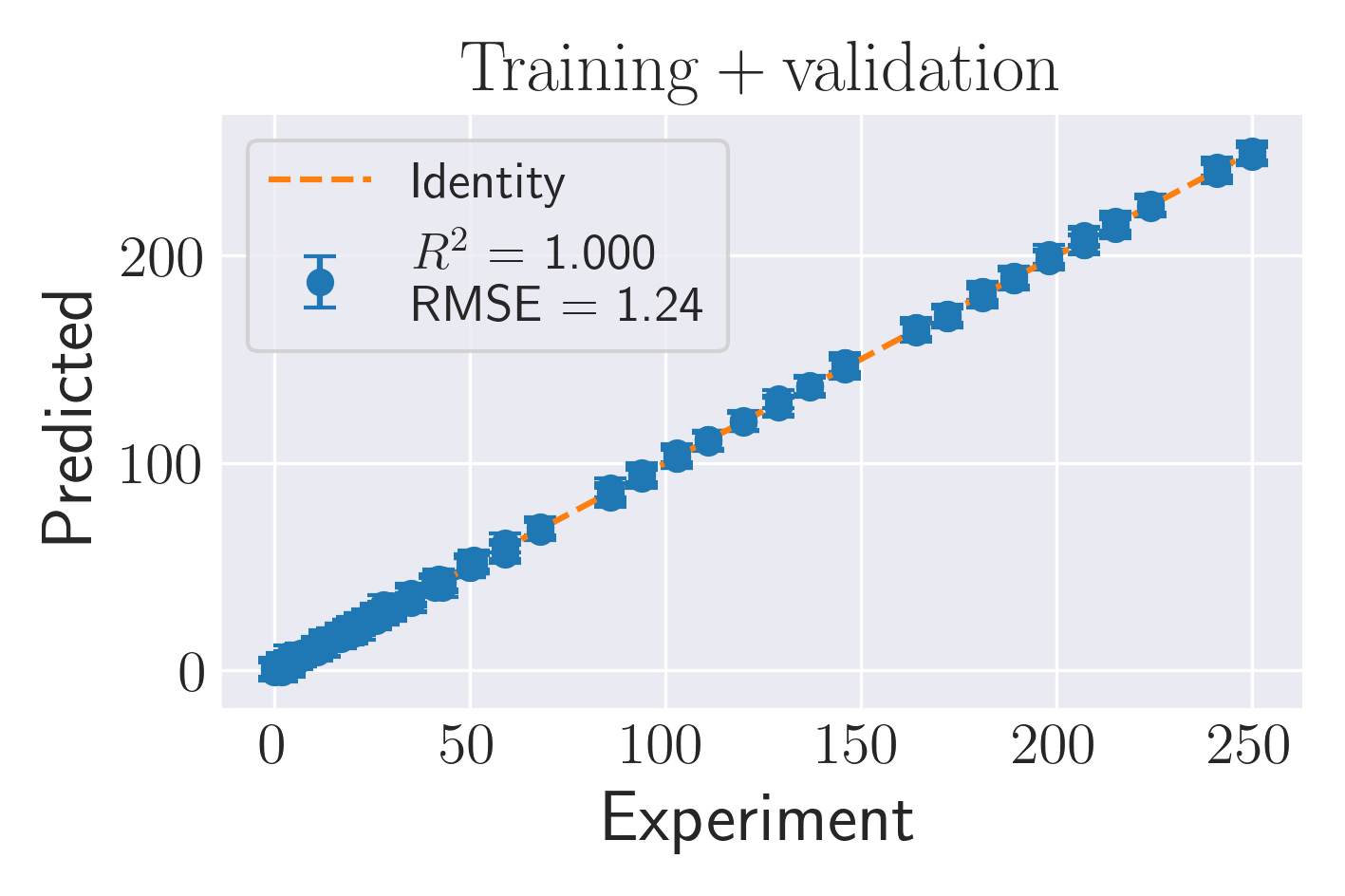}
\end{subfigure}%
\begin{subfigure}{0.23\textwidth}
    \centering
    \includegraphics[width=\linewidth]
    {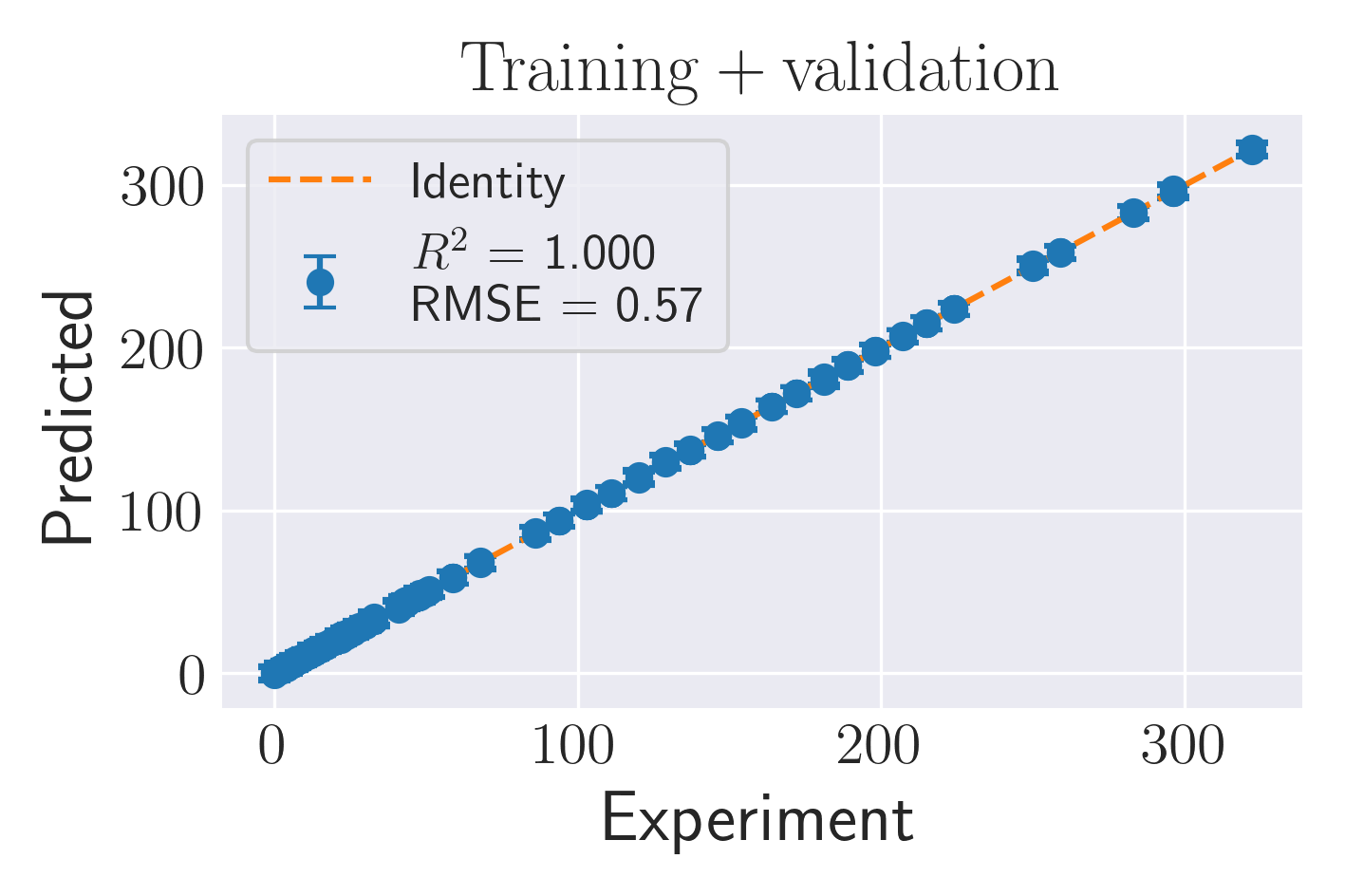}
\end{subfigure}

% =====================================================================
% FULL-FEATURE MODEL: TEST
% =====================================================================
\begin{subfigure}{0.23\textwidth}
    \centering
    \includegraphics[width=\linewidth]
    {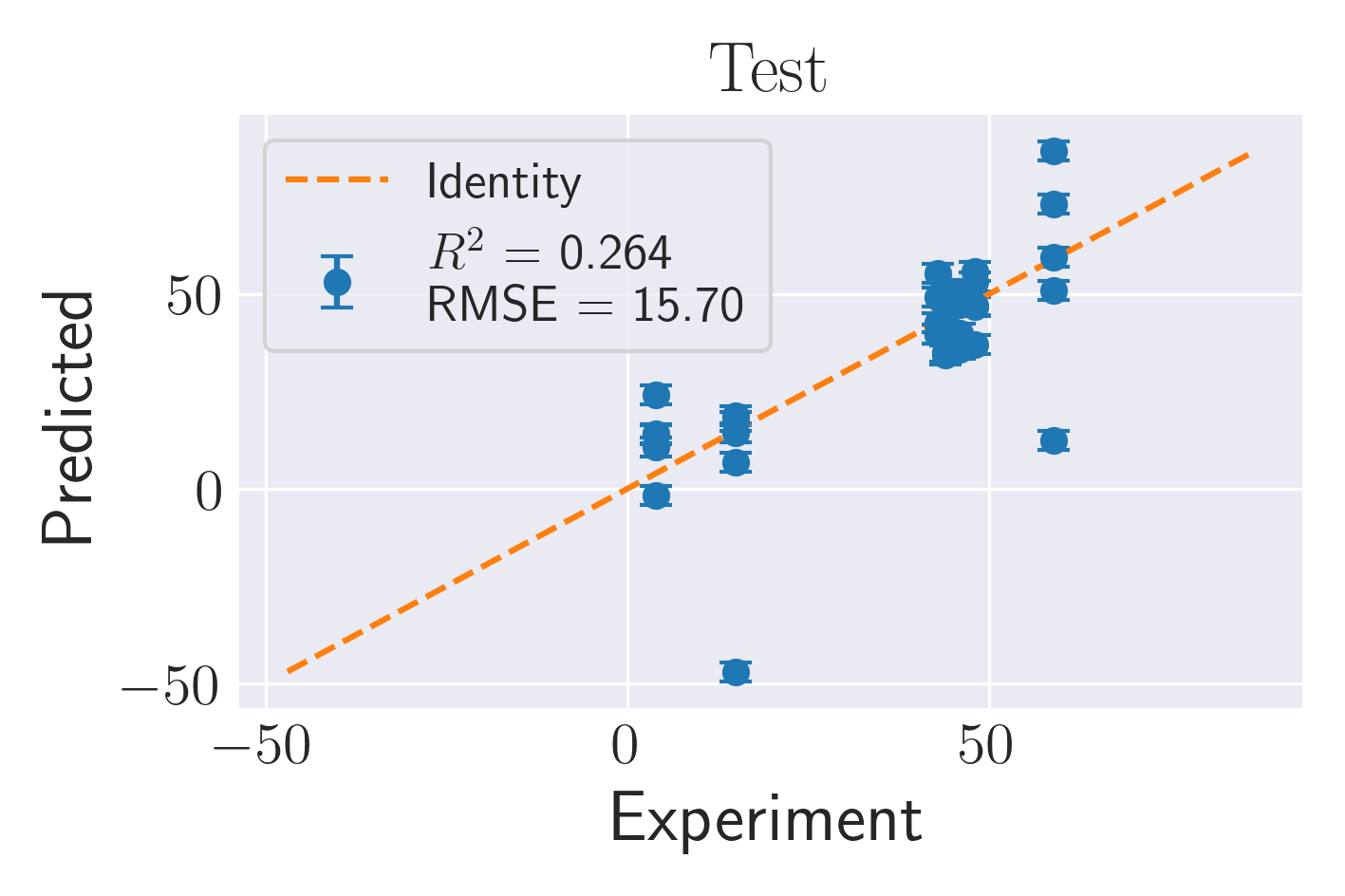}
\end{subfigure}%
\begin{subfigure}{0.23\textwidth}
    \centering
    \includegraphics[width=\linewidth]
    {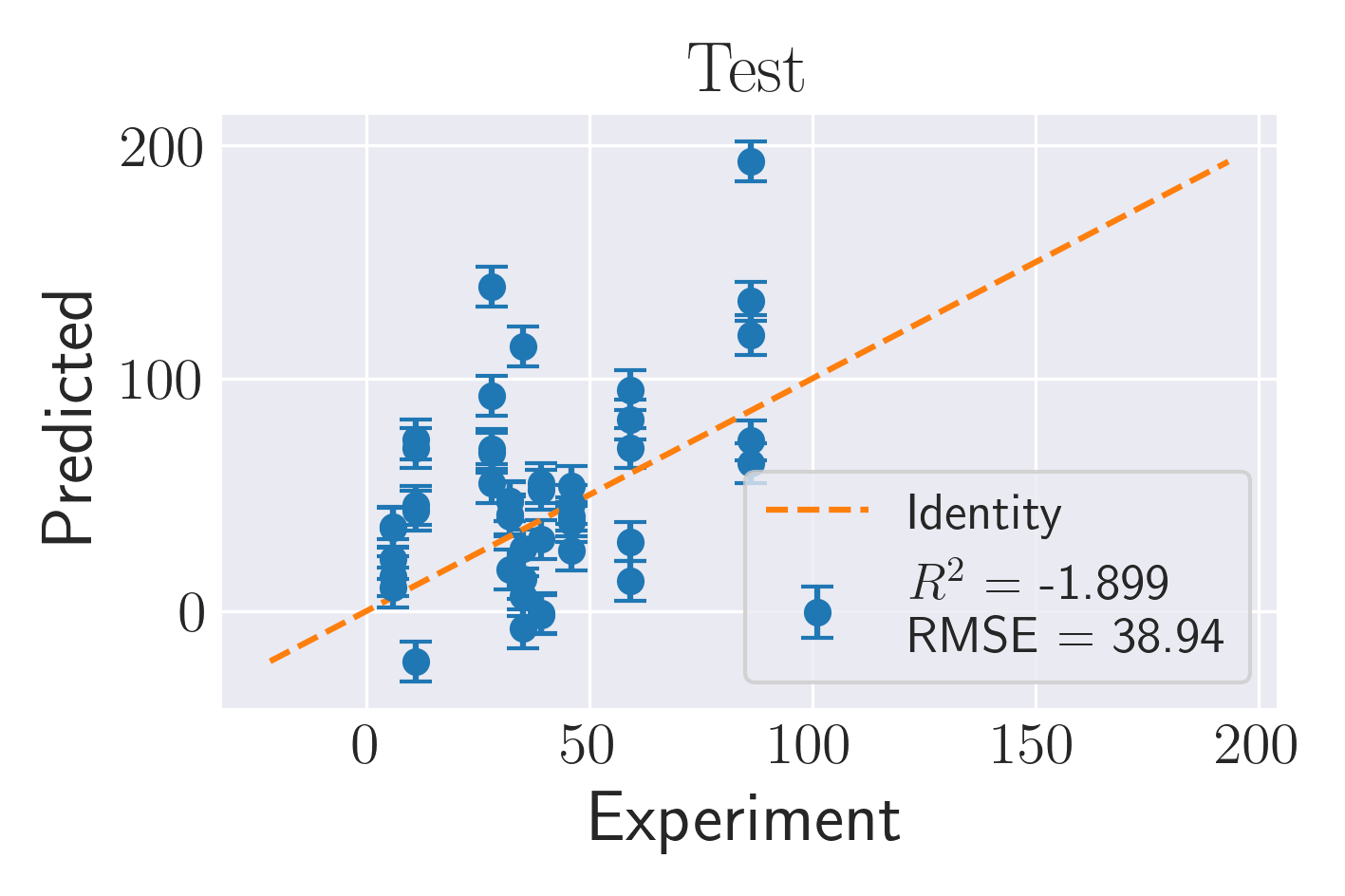}
\end{subfigure}%
\begin{subfigure}{0.23\textwidth}
    \centering
    \includegraphics[width=\linewidth]
    {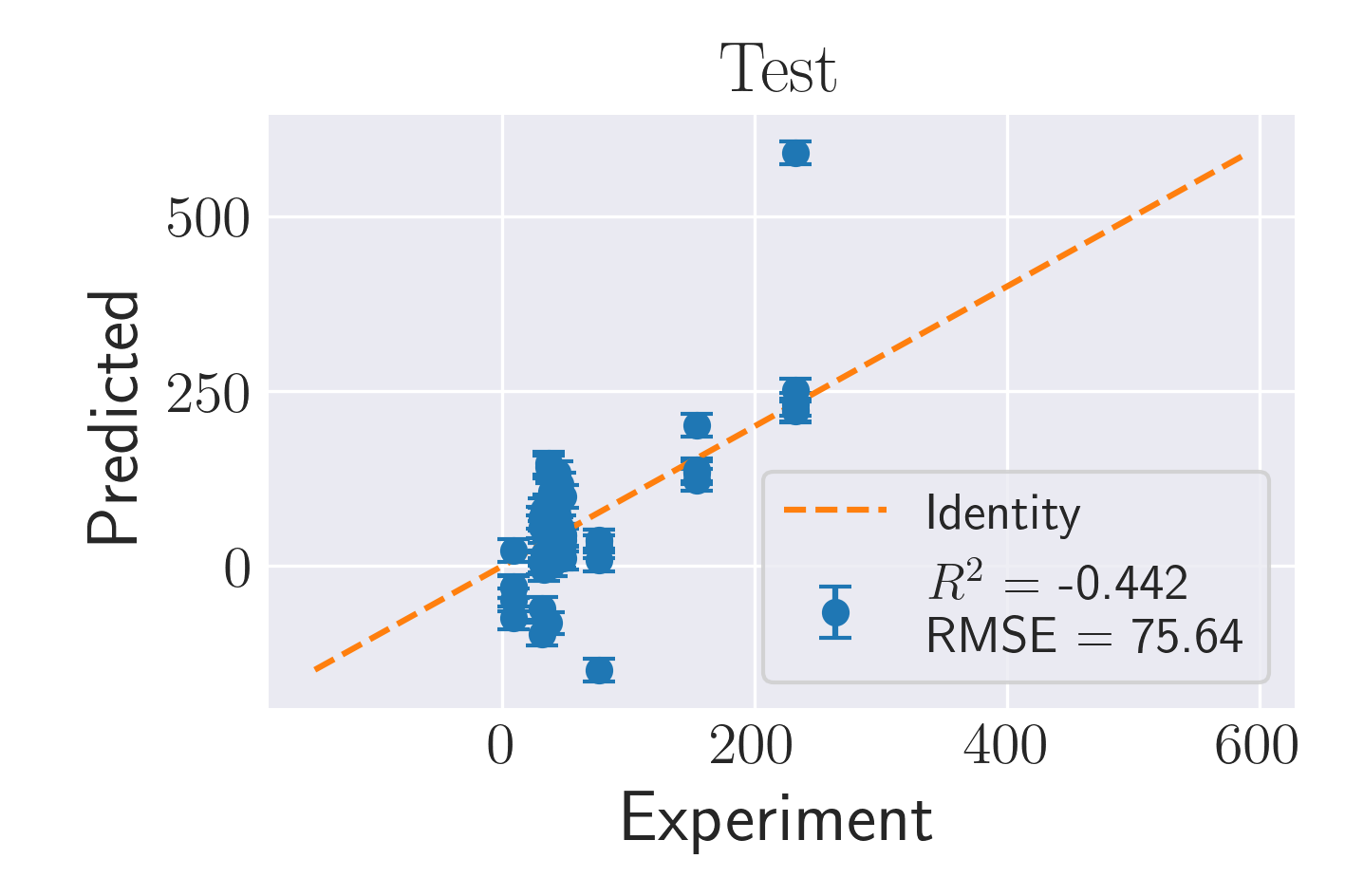}
\end{subfigure}%
\begin{subfigure}{0.23\textwidth}
    \centering
    \includegraphics[width=\linewidth]
    {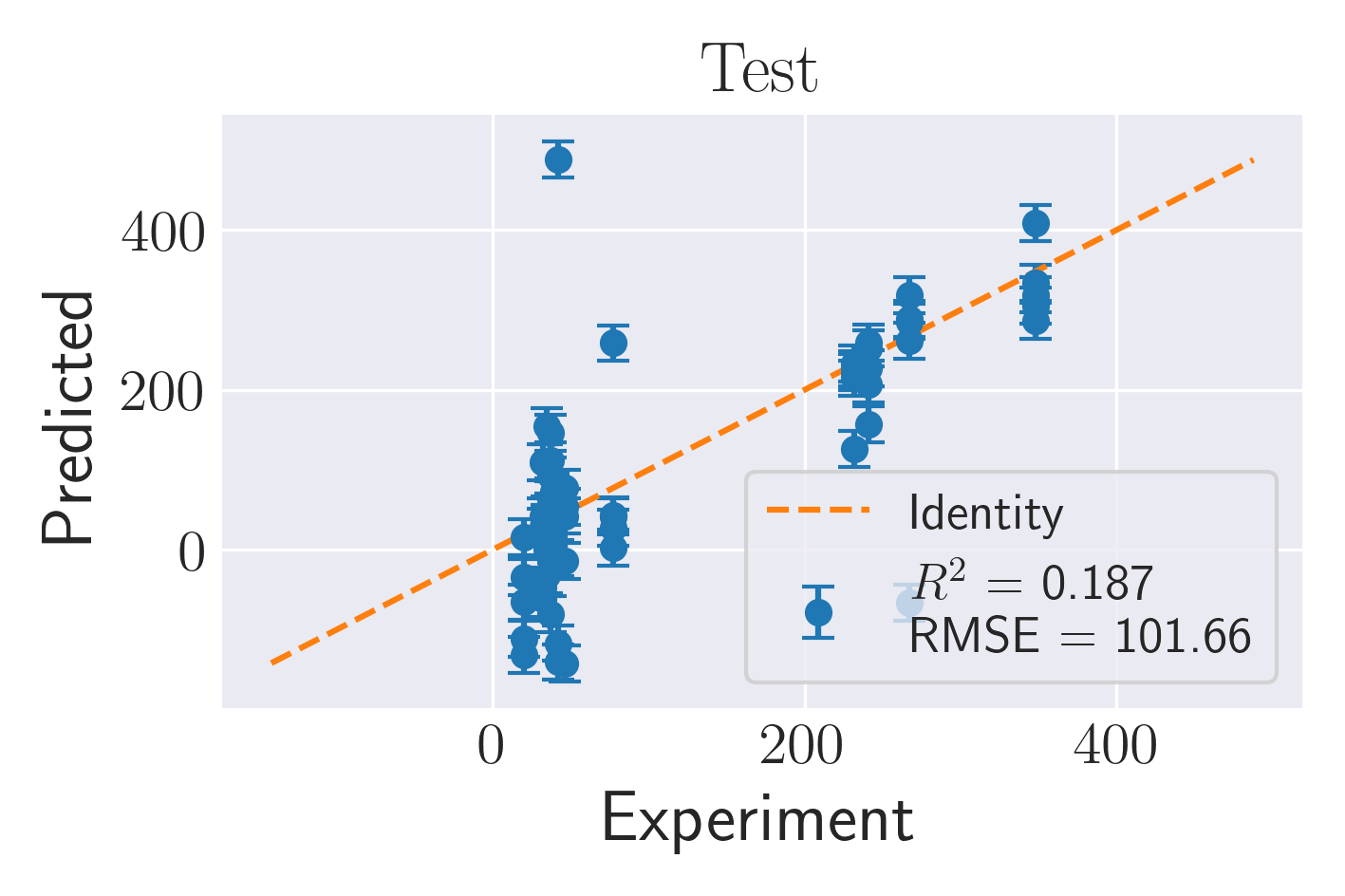}
\end{subfigure}

\caption{Parity plots for the soft sensor across progressively broader nominal glucose
concentration ranges: A) 0--75~g/L; B) 0--150~g/L;
C) 0--250~g/L; and D) 0--350~g/L. For each concentration
range, the first two rows show the training and held-out validation
performance, respectively, of the feature-region model selected through
the validation-based model-selection procedure. The selected model is
then refitted using the combined training and validation sets, with the
third and fourth rows showing its performance on the combined
training-validation data and the held-out test set, respectively.
The fifth and sixth rows show the corresponding training-validation
and test performance of a full-feature soft-sensor benchmark using the
complete set of voltammetric read-out channels. Error bars represent one posterior standard deviation, with the fitted observation-noise contribution excluded.}

\label{fig:model_selection_performance}
\end{figure*}

As shown in column A of Fig.~\ref{fig:model_selection_performance}, the selected model achieved a validation $R^2$ of 0.977 and an RMSE of 1.90~g/L during model selection. After refitting using the combined training and validation sets, it achieved an $R^2$ of 0.999 and an RMSE of 0.39~g/L on those data. Evaluation on the held-out test concentration levels gave an $R^2$ of 0.990 and an RMSE of 1.84~g/L. By comparison, the full-feature GPR benchmark achieved an $R^2$ of 1.000 and an RMSE of 0.28~g/L on the combined training and validation data, but its test performance deteriorated to an $R^2$ of 0.264 and an RMSE of 15.70~g/L, signalling overfitting. The selected feature-region model therefore reduced the test RMSE by 88.3\%. For this concentration range, the results indicate that feature-region selection
substantially improved generalisation relative to the full-feature benchmark.

\subsection{Automated soft-sensor development for the 0--150 g/L range}

The nominal glucose concentration range was next expanded to 0--150~g/L to assess the framework over a broader sensing window. Column B of Fig.~\ref{fig:scores} presents the squared Spearman's rank correlation coefficient, the measurement-quality metric, and the resulting composite information fingerprint for the oxidation and reduction channels. The retained candidate intervals are also shown. In total, the information fingerprints yielded four candidate regions: three from the oxidation branch and one from the reduction branch. Evaluating all non-empty combinations of these regions resulted in $2^4-1=15$ candidate GPR models. The range-specific training subset therefore yielded one fewer candidate region than that used for the 0--75~g/L analysis.

Validation-based ranking selected a single oxidation region spanning 0.60 to 0.71~V, as summarised in Table~\ref{tab:final_models}. This region contains 12 of the original 400 measurement channels, representing a 97.0\% reduction in input dimensionality. Notably, this region had the highest average fingerprint score among the retained candidates but did not coincide with the dominant current-magnitude peaks visible in the original voltammograms. This result further demonstrates that predictive relevance is not determined solely by the magnitude of the raw current response.

As shown in column B of Fig.~\ref{fig:model_selection_performance}, the selected model achieved a validation $R^2$ of 0.924 and an RMSE of 7.01~g/L during model selection. After refitting using the combined training and validation sets, it achieved an $R^2$ of 0.999 and an RMSE of 1.23~g/L on those data. Evaluation on the held-out test concentration levels gave an $R^2$ of 0.989 and an RMSE of 2.38~g/L. By comparison, the full-feature GPR benchmark achieved an $R^2$ of 1.000 and an RMSE of 0.52~g/L on the combined training and validation data, but its test performance deteriorated to an $R^2$ of $-1.899$ and an RMSE of 38.94~g/L. The selected feature-region model therefore reduced the test RMSE by 93.9\%. For this broader concentration range, the results again indicate that feature-region selection substantially improved generalisation relative to the full-feature benchmark.

\subsection{Automated soft-sensor development for the 0--250 g/L range}

The nominal glucose concentration range was expanded further to 0--250~g/L. Column C of Fig.~\ref{fig:scores} presents the squared Spearman's rank correlation coefficient, the measurement-quality metric, and the resulting composite information fingerprint for the
oxidation and reduction channels. The retained candidate intervals are also shown. In total, the information fingerprints yielded six candidate regions: three from the oxidation branch and three from the reduction branch. Evaluating all non-empty combinations of these
regions resulted in $2^6-1=63$ candidate GPR models. Compared with the five and four candidate regions obtained for the 0--75 and 0--150~g/L ranges, respectively, this result further shows that the data-derived candidate set changes as the concentration window is
broadened.

Validation-based ranking selected two regions: an oxidation region spanning approximately 0.66 to 0.74~V and a reduction region spanning approximately 0.08 to 0.11~V, as summarised in Table~\ref{tab:final_models}. Together, these regions contain 11 of the original 400 measurement channels, representing a 97.3\% reduction in input dimensionality. Notably, the oxidation region overlaps with the region selected for the 0--150~g/L range. The reduction region provides complementary predictive information. This result further illustrates how validation-based model selection identifies jointly predictive interval combinations while retaining a parsimonious input representation.

As shown in column C of Fig.~\ref{fig:model_selection_performance}, the selected model achieved a validation $R^2$ of 0.998 and an RMSE of 4.03~g/L during model selection. After refitting using the combined training and validation sets, it achieved an $R^2$ of 0.998 and an RMSE of 3.53~g/L on those data. Evaluation on the held-out test concentration levels gave an $R^2$ of 0.996 and an RMSE of 4.09~g/L. By comparison, the full-feature GPR benchmark achieved an $R^2$ of 1.000 and an RMSE of 1.24~g/L on the combined training and validation data, but its test performance deteriorated to an $R^2$ of $-0.442$ and an RMSE of 75.64~g/L. The selected feature-region model therefore reduced the test RMSE by 94.6\%. For this broader concentration range, the results further indicate that feature-region selection substantially improved generalisation relative to the full-feature benchmark.

\subsection{Automated soft-sensor development for the 0--350 g/L range}

Finally, the framework was evaluated over the broadest nominal glucose
concentration range of 0--350~g/L. Column D of Fig.~\ref{fig:scores} presents the squared Spearman's rank correlation coefficient, the measurement-quality metric, and the resulting composite information fingerprint for the oxidation and reduction channels. The retained candidate intervals are also shown. In total, the information fingerprints yielded six candidate regions: three from the oxidation branch and three from the reduction branch. These regions closely correspond to those obtained for the 0--250~g/L range, indicating stability of the candidate locations across these two range-specific training subsets. Evaluating all non-empty combinations of these regions resulted in $2^6-1=63$ candidate GPR models.

Validation-based ranking selected three regions: oxidation regions with peak-width boundaries of approximately 0.66 to 0.74~V and 0.27 to 0.36~V, and a reduction region spanning approximately 0.35 to 0.41~V, as summarised in Table~\ref{tab:final_models}. Together, these regions contain 22 of the original 400 measurement channels, representing a 94.5\% reduction in input dimensionality. Notably, the high-potential oxidation region broadly overlapped in all four investigated concentration ranges. By contrast, the selected reduction regions varied between concentration ranges, suggesting a more
concentration-window-dependent complementary contribution. 

As shown in column D of Fig.~\ref{fig:model_selection_performance}, the
selected model achieved a validation $R^2$ of 0.994 and an RMSE of 7.93~g/L during model selection. After refitting using the combined training and validation sets, it achieved an $R^2$ of 1.000 and an RMSE of 0.40~g/L on those data. Evaluation on the held-out test concentration levels gave an $R^2$ of 0.992 and an RMSE of 9.98~g/L. By comparison, the full-feature GPR benchmark achieved an $R^2$ of 1.000 and an RMSE of 0.57~g/L on the combined training and validation data, but its test performance deteriorated to an $R^2$ of 0.187 and an RMSE of
101.66~g/L. The selected feature-region model therefore reduced the test RMSE by 90.2\%. As in all previous cases, feature-region selection substantially improved generalisation relative to the full-feature benchmark.

\section{Conclusions}

In this work, we introduced an automated feature-region selection framework for developing soft sensors from spectral-like measurements, which are widely generated by common PAC and PAT methods. The ordered, high-dimensional feature spaces associated with these measurements can increase data requirements and susceptibility to overfitting. The framework constructs an information fingerprint by combining channel-level target correlation with a replicate-based signal-to-noise indicator, derives contiguous candidate intervals from this latent profile, and evaluates their combinations according to predictive performance and parsimony. Candidate interval generation is independent of the downstream regression architecture and can therefore be coupled, in principle, with different regression models. GPR was used in this study because it can represent nonlinear relationships, account for observation noise, and provide posterior uncertainty estimates while remaining suitable for small to medium-sized experimental datasets.

The framework was demonstrated using cyclic voltammetric measurements of glucose acquired with an Au electrode. Across the four investigated concentration ranges, the selected models used between 11 and 42 of the original 400 measurement channels, representing reductions in input dimensionality of 89.5--97.3\%. Relative to the corresponding full-feature GPR benchmarks, the selected models reduced the held-out test RMSE by 88.3--94.6\%. The selected regions were not consistently apparent from the magnitude of the raw measurements, demonstrating the value of combining target correlation, measurement quality, and model-based evaluation when identifying informative feature regions. The near-perfect fits obtained by the full-feature models on the combined training and validation data, followed by their markedly poorer test performance, are consistent with overfitting in the 400-channel representation. Feature-region selection retained strong calibration performance while substantially improving prediction at held-out concentration levels. These results demonstrate that compact representations formed from selected contiguous regions of the original signal can improve soft-sensor generalisation.

The glucose measurements in PBS provided a controlled proof of concept for the proposed framework. A deployment-ready glucose sensor would require future work, including implementation and robustness checks using online or inline measurements acquired from process broths and other process-specific matrices. That said, the principal contribution of this paper remains the automated feature-selection framework for facilitating soft-sensor development in the process industries.
\newline

% \newline
\noindent \textbf{CRediT authorship contribution statement} \newline
\textbf{Sebastián Espinel-Ríos}: Conceptualization, Methodology, Software, Formal Analysis, Writing - Original Draft, Writing - Review \& Editing, Visualization.
\textbf{Wenchao Duan}: Conceptualization, Investigation, Validation,  Writing - Original Draft, Writing - Review \& Editing, Visualization.
\newline

\noindent \textbf{Declaration of interests} \newline
The authors declare that they have no conflict of interest. \newline

\noindent \textbf{Data statement} \newline
Data will be made available upon reasonable request. \newline

\bibliographystyle{elsarticle-num} 
\bibliography{bibliography}

\end{document}